\documentclass[12pt]{article}
\usepackage{amsfonts}
\usepackage{amsmath,amssymb} 
\usepackage{graphicx} 
\usepackage[dvipsnames,x11names]{xcolor} 
\usepackage{hyperref} 
\hypersetup{bookmarksopen=true, bookmarksnumbered=true,
  pdfstartview={FitH}, pdfborder={0 0 0}, colorlinks=true,
  linkcolor=red, linktocpage=true, citecolor=blue, urlcolor=cyan,
 unicode=true, pdftitle={interpolating function}, pdfauthor={DJ},
pdfsubject={interpolating function},
 pdfkeywords={interpolating function}}
\usepackage{geometry} 
\usepackage{fancyhdr} 
\usepackage[nottoc,notlof,notlot]{tocbibind} 
\usepackage[titles]{tocloft} 
\usepackage{epsfig}
\usepackage{latexsym}
\usepackage{color}

\numberwithin{equation}{section} 

\newcommand{\eq} {equation}
\newcommand{\eqa} {eqnarray}
\newcommand{\NN} {\mbox {$\nonumber$}}

\begin{document}
\title{\LARGE\textbf{Interpolating function and Stokes Phenomena}}
\bigskip
\author{\textbf{ Masazumi
    Honda}\footnote{\href{mailto:masazumihonda@hri.res.in}{masazumihonda@hri.res.in}}\ \
  and\ \textbf{Dileep
    P. Jatkar}\footnote{\href{mailto:dileep@hri.res.in}{dileep@hri.res.in}}
  \bigskip\bigskip\\
   Harish-Chandra Research Institute\\
    Chhatnag Road, Jhunsi\\
 Allahabad 211019, India
  }
\date{}
\maketitle
\thispagestyle{fancy}
\vspace{3mm}
\begin{abstract}
  When we have two expansions of physical quantity around two
  different points in parameter space, we can usually construct a
  family of functions, which interpolates the both expansions.  In
  this paper we study analytic structures of such interpolating
  functions and discuss their physical implications.  We propose that
  the analytic structures of the interpolating functions provide
  information on analytic property and Stokes phenomena of the
  physical quantity, which we approximate by the interpolating
  functions.  We explicitly check our proposal for partition functions
  of zero-dimensional $\varphi^4$ theory and Sine-Gordon model.  In
  the zero dimensional Sine-Gordon model, we compare our result with a
  recent result from resurgence analysis.  We also comment on
  construction of interpolating function in Borel plane.
\end{abstract}
\vfill
\noindent HRI/ST/1501 
\newpage
\tableofcontents
\setcounter{page}{1}
\pagestyle{plain}
\section{Introduction}
In most non-exactly solvable problems, one tries to look for a small
parameter or a large parameter so that we can set up a perturbative
expansion.  This perturbation series is typically an asymptotic series
which is divergent and has zero radius of convergence.  Often the
perturbation series takes different forms depending on the argument of
its expansion parameter.  This behavior is known as {\it the Stokes
  phenomenon} \cite{Stokes}.  For example, if we consider a physical
quantity having an integral representation as in quantum field theory
and string theory, then the integrand usually has multiple saddle
points and its perturbative expansion with the parameter $g$ typically
takes the form
\begin{\eq}
 \sum_k a_{0,k} g^k +e^{-S_1 (g)} \sum_k a_{1,k} g^k +e^{-S_2 (g)} \sum_k a_{2,k} g^k +\cdots ,
\label{eq:series}
\end{\eq}
where $S_i (g) $ is the ``action" evaluated at each saddle point and
each sum denotes the small-$g$ expansion around that saddle point.  In
standard situations, the weight $e^{-S_i (g)}$ is exponentially
suppressed for real positive $g$ and vanishes as approaching
$g\rightarrow +0$.  In other words all the terms except the first one
in \eqref{eq:series} describe non-perturbative effects.  However, if
we change ${\rm arg}(g)$, then $S_i (g)$ might have negative real
part.  For this case, the weight $e^{-S_i (g)}$ is not exponentially
suppressed but exponentially growing.  This type of transition happens
when we cross an anti-Stokes line given by ${\rm Re}(S_i )=0$.  On the
other hand, the coefficients $a_{i,k}$ themselves might jump as we
change ${\rm arg}(g)$.  This happens when we cross a Stokes line
defined by ${\rm Im}(S_i )={\rm Im}(S_j )$.  Thus Stokes phenomena has deep
connections to non-perturbative effects.

In most practical situations,
we can access to only first few terms of a perturbation series around
single saddle point for particular argument of expansion parameter.
For this case, it is hard to find when Stokes phenomena occur and when
contributions from other saddle points become important.
In this paper we develop a tool to study Stokes phenomena
in somewhat special situations.

Sometimes we can have two perturbative expansions of physical
quantity around two different points in parameter space, for example,
in theory with S-duality, field theory with gravity dual,
lattice gauge theory with weak and strong coupling expansions,
statistical system with high and low temperature expansions, and so
on.  One can then construct functions, which interpolate these two
expansions.  The most standard approach to this is (two-point) Pad\'e
approximant, which is a rational function having the two expansions up
to some orders.  Recently Sen has considered another type of
interpolating function, which has the form of a Fractional Power of
Polynomial (FPP) \cite{Sen:2013oza}.  After a while, one of us has
constructed a more general class of interpolating functions described
by Fractional Powers of Rational function (FPR) \cite{Honda:2014bza},
which includes the Pad\'e approximant and FPP as special cases.  It
has turned out that these interpolating functions usually provide
better approximations than each perturbative expansion in intermediate
regime of the parameter, see
\cite{Sen:2013oza,Honda:2014bza,Asnin:2007rw,Banks:2013nga,Pius:2013tla,Alday:2013bha}
for various applications\footnote{
  There is also another type of interpolating functions
  \cite{Kleinert:2001ax}, which is not special case of the FPR.  This
  has been applied to $O(N)$ non-linear sigma model.  }.  Although
these are quite nice as first attempts, properties of interpolating
functions themselves have not been extensively studied yet.

In this paper we propose new properties of the interpolating
functions.  We focus on analytic structure of the interpolating
functions\footnote{ Note that our interpolating function approach is
  different from resurgence approach, which has been recently studied
  in a series of works
  \cite{Argyres:2012ka,Argyres:2012vv,Dunne:2012zk,Dunne:2012ae,Dunne:2013ada,Schiappa:2013opa,Cherman:2013yfa,Aniceto:2013fka,Basar:2013eka,Dunne:2014bca,Cherman:2014ofa,Cherman:2014xia,Aniceto:2014hoa,Couso-Santamaria:2015wga,Basar:2015xna}.
  While we use single types of weak and strong coupling expansions as
  input data, the resurgence approach uses weak coupling expansions
  around multiple saddle points and does not use strong coupling
  expansion.  }  by treating them as complex functions and discuss
their physical implications.  The FPR is some power of a rational
function and the rational function has poles and zeros.  When the
power is not integer, then the poles and zeros of the rational
function give rise to branch cuts of the FPR.  Here we propose that
the branch cuts of the FPR encode information about the analytic
property and Stokes phenomena of the physical quantity, which we try
to approximate.  More concretely we propose that each branch cut of
the FPR has the following possible interpretations.
\begin{enumerate}
\item The branch cut is particular to the FPR and the artifact of
  the approximation.  Namely, this type of branch cut is not useful to
  extract physical information.

\item The physical quantity, which we approximate by the FPR, has a
  branch cut near from the branch cut of the FPR.  Namely, the branch
  cut of the FPR well approximates the ``true" branch cut of the
  physical quantity.

\item Near the branch cut, one of perturbation series of the physical
  quantity changes its dominant part.  This case is further separated into the following two
  possibilities.
\begin{enumerate}
\item We have an {\it anti-Stokes line} of the perturbative expansion
  near the branch cut.  Namely, although the perturbative series
  itself {\it does not change} its own form, contributions from other
  saddle points become dominant across the line.  This possibility
  likely occurs for first branch cut measured from a specific axis
  where we construct interpolating functions.

\item The perturbative series itself {\it does change} the form.
  Namely, we have a {\it Stokes line} near the branch cut, whose
  diagonal multiplier is different from 1.  When we have a Stokes line
  across which sub-dominant parts of the perturbative series change,
  the FPR cannot detect this type of Stokes line.
\end{enumerate}
\end{enumerate}

We explicitly check our proposal in two examples: partition functions
of the $\varphi^4$ theory and the Sine-Gordon model in zero
dimensions.  
Similar features seem to appear also in other examples such as BPS Wilson loop
in 4d $\mathcal{N}=4$ Super Yang-Mills theory, energy spectrum in 1d
anharmonic oscillator etc \cite{progress}.  
We expect that our result is applicable in more practical
problems, where we do not know exact solutions.  
One possible utility of
such an analysis is that we can anticipate analytic property and
Stokes phenomena of physical quantity by looking
at analytic structures of interpolating functions.  

This paper is organized as follows.  In section
\ref{sec:interpolating} we introduce our interpolating functions
described by the fractional power of rational functions (FPRs).  In
section \ref{sec:phi4} we study interpolating problem in the 0d
$\varphi^4$ theory in great detail.  We look at the analytic property
of the interpolating function as a complex function and propose its
physical interpretation.  In section \ref{sec:SG} we analyze the 0d
Sine-Gordon model and check that our proposal is true also for this
model. 
Section \ref{sec:conclusion} is devoted to
conclusion and discussions.
In appendix \ref{app:resurgence}
we compare
our interpolating function with a recent result from resurgence
analysis \cite{Cherman:2014xia} in the 0d Sine-Gordon model.  
Our result implies that the FPR and
resurgence play complementary role with each other.  
In appendix \ref{sec:Borel} we explain an attempt to construct
interpolating function in Borel plane and test its utility in the 0d
Sine-Gordon model. 
In appendix \ref{app:explicit} we write down explicit forms for interpolating
functions used in the main text.

\section{Interpolating function}
\label{sec:interpolating}
We introduce the interpolating functions in this section,
which is essentially a review of
\cite{Honda:2014bza}.  Suppose that we wish to determine a function
$F(g)$, which has\footnote{ More generally we might have perturbative
  expansions around $g=g_1$ and $g=g_2$ with $g_2 > g_1$ and would
  like to construct their interpolating functions.  However, if we
  change the variable as $x=(g-g_1)/(g_2 -g)$, then this problem is
  reduced to interpolating problem of small-$x$ and large-$x$
  expansions.  Thus our setup does not lose
  generality in this sense.  } the small-$g$ expansion
$F_s^{(N_s)}(g)$ and large-$g$ expansion $F_l^{(N_l )}(g)$ taking the forms
\begin{equation}
  F_s^{(N_s )}(g) = g^a \sum_{k=0}^{N_s} s_k g^k ,\quad
  F_l^{(N_l )}(g) = g^b \sum_{k=0}^{N_l} l_k g^{-k} .
\label{eq:asymptotics}
\end{equation}
We can then naively expect that these expansions approximate $F(g)$ as
\begin{\eq}
F(g) 
= F_s^{(N_s )}(g) +\mathcal{O}(g^{a+N_s -1})
= F_l^{(N_l )}(g) +\mathcal{O}(g^{b-N_l  -1}) .
\end{\eq}
Although this seems to be a somewhat limited case, this situation
includes a large class of physical problems, e.g., 
theory with S-duality, field theory with gravity dual, lattice gauge
theory with weak and strong coupling expansions, statistical system
with high and low temperature expansions, and etc.

In terms of the two expansions,
one can construct the following function \cite{Honda:2014bza}
\begin{equation}
F_{m,n}^{(\alpha )} (g)
= s_0 g^a \Biggl[ \frac{ 1 +\sum_{k=1}^p c_k g^k}{1 +\sum_{k=1}^q d_k
  g^k }  \Biggr]^\alpha ,
\label{eq:FPR}
\end{equation}
where 
\begin{equation}\label{eq:2}
p = \frac{1}{2} \left( m+n+1 -\frac{a-b}{\alpha} \right) ,\quad
q = \frac{1}{2} \left( m+n+1 +\frac{a-b}{\alpha} \right) .
\end{equation}
The coefficients $c_k$ and $d_k$ are determined such that series
expansions of $F_{m,n}^{(\alpha )} (g)$ around $g=0$ and $g=\infty$
reproduce the small-$g$ and large-$g$ expansions
(\ref{eq:asymptotics}) of $F(g)$ up to $\mathcal{O}(g^{a+m+1})$ and
$\mathcal{O}(g^{b-n-1})$, respectively.  Due to this property, the
function $F_{m,n}^{(\alpha )} (g)$ interpolates the small-$g$ and
large-$g$ expansions up to these orders.  Since the interpolating
function is usually\footnote{ When $a-b$ is irrational number, the
  power $\alpha$ is irrational number.  } described by the Fractional
Power of Rational function, we call this type of the interpolating
function ``FPR".  Note that the rational function inside the square
bracket in (\ref{eq:FPR}) is a ratio of polynomials, {\em i.e.},
\begin{equation}
p,q \in \mathbb{Z}_{\geq 0} ,
\end{equation}
which leads the following condition
\begin{equation}
\alpha = \left\{ \begin{matrix}
\frac{a-b}{2\ell +1}  & {\rm for} & m+n:{\rm even} \cr
\frac{a-b}{2\ell}  & {\rm for} & m+n:{\rm odd} \end{matrix} \right. ,\quad
{\rm with}\ \ell \in\mathbb{Z} .
\end{equation}

It is now easy to see that
the FPR includes the Pad\'e approximant and the Fractional Power of Polynomial (FPP) constructed in \cite{Sen:2013oza} 
as special cases.  
If we take $2\ell+1=a-b$ for $a-b \in \mathbb{Z}$ and $m+n$
to be even, then this becomes the Pad\'e approximant 
while if we take $2\ell+1=m+n+1$ ($2\ell=m+n+1 $) for even (odd) $m+n$
then we get the FPP.  
Therefore we below refer to also the Pad\'e and FPP as FPR.

\section{Partition function of zero-dimensional $\varphi^4$ theory}
\label{sec:phi4}
\begin{figure}[t]
\begin{center}
\includegraphics[width=7.4cm]{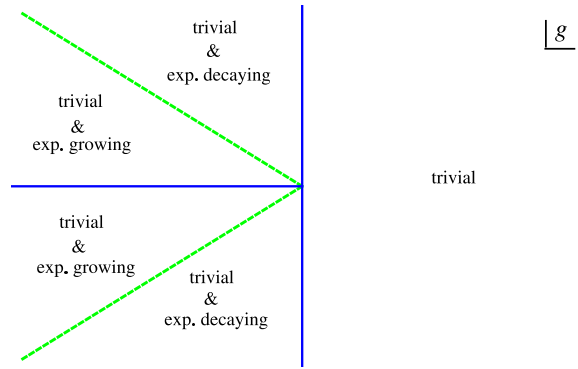}
\end{center}
\caption{Summary of Stokes phenomena for the small-$g$ expansion
in the partition function of the 0d $\varphi^4$ theory.
The blue solid lines denote the Stokes lines
while the green dashed lines denote the anti-Stokes lines.}
\label{fig:phi4_0d_stokes}
\end{figure}
In this section we study interpolation problem for the partition function
of the 0d $\varphi^4$ theory.  Although this example has been already
studied well in \cite{Sen:2013oza,Honda:2014bza} for real
non-negative coupling constant, here we consider general complex
coupling.  As mentioned above, we study analytic properties of
interpolating functions and their physical implications.
Let us consider the integral
\begin{equation}
F(g) = \frac{1}{\sqrt{g}}\int_{-\infty}^\infty dx \ e^{-\frac{x^2}{2g} -x^4} ,
\end{equation}
which can be exactly performed as
\begin{equation}
F(g)
=\left\{\ \begin{matrix}
\frac{\pi  e^{\frac{1}{32 g^2}} }{4 g}
 \left[I_{-\frac{1}{4}}\left(\frac{1}{32 g^2}\right) -I_{\frac{1}{4}}\left(\frac{1}{32 g^2}\right)\right]
& {\rm for}\ {\rm Re}(g)>0 \cr \cr
\frac{\pi  e^{\frac{1}{32 g^2}} }{4 \sqrt{-g^2}}
 \left[I_{-\frac{1}{4}}\left(\frac{1}{32 g^2}\right)+I_{\frac{1}{4}}\left(\frac{1}{32 g^2}\right)\right]
& {\rm for}\ {\rm Re}(g)\leq 0 \cr
\end{matrix}\right. ,
\end{equation} 
where $I(z)$ is the modified Bessel function of the first kind.
The small-$g$ expansion of $F(g)$ depends on the argument of
$g$, and its dependence is given by
\begin{equation}
\label{eq:8}
F(g)
=\left\{ \begin{matrix}
\sqrt{2 \pi }-3 \sqrt{2 \pi } g^2+105 \sqrt{\frac{\pi }{2}} g^4+\mathcal{O}(g^6 )
& {\rm for}\ {\rm arg}(g)\in ( -\frac{\pi}{2},\frac{\pi}{2}) \cr
\sqrt{\frac{\pi}{2} } \left(105 g^4-6 g^2+2 +\mathcal{O}(g^6 ) \right) -\sqrt{\pi } i e^{\frac{1}{16 g^2}} \left(105 g^4+6 g^2+2 +\mathcal{O}(g^6 ) \right) 
& {\rm for}\ {\rm arg}(g)\in ( \frac{\pi}{2},\pi) \cr
\sqrt{\frac{\pi}{2} } \left(105 g^4-6 g^2+2 +\mathcal{O}(g^6 ) \right) +\sqrt{\pi } i e^{\frac{1}{16 g^2}} \left(105 g^4+6 g^2+2 +\mathcal{O}(g^6 ) \right) 
& {\rm for}\ {\rm arg}(g)\in ( -\pi ,-\frac{\pi}{2}) \cr
\end{matrix} \right. .
\end{equation}
This dependence on ${\rm arg}(g)$ clearly reflects the fact that we
have Stokes lines oriented along ${\rm arg}(g)=\pm\pi/2$ and $\pi$,
across which the form of the small-$g$ expansion changes.  Notice that
across the ray $|{\rm arg}(g)|> 3\pi/4$, the terms involving the
exponential factor (in line 2 and 3 in eq.~(\ref{eq:8})) are dominant
compared to those without the exponential factor.  This indicates
that we have the anti-Stokes lines oriented along ${\rm arg}(g)=\pm 3\pi/4$.

It is easy to understand this behavior from the standard saddle point
analysis for $|g|\ll 1$.  Saddle points $x_\ast$ of the integration
are given by
\begin{\eq}
x_\ast = 0, x_\pm ,\quad {\rm with}\ x_\pm = \pm \frac{i}{2\sqrt{g}} .
\end{\eq}
Then, the ``action" $S(x)=\frac{x^2}{2g} +x^4$ at the saddle points $x_\ast$ takes the values
\begin{\eq}
S(x_\ast =0 ) =0,\quad S(x_\ast =x_\pm ) =-\frac{1}{16g^2} .
\end{\eq}
For $|{\rm arg}(g)|<\pi /2$, we can pick up only the trivial saddle
point $x_\ast =0$ by deforming the original integral contour
$(-\infty ,\infty )$ to a steepest descent contour, while we can
pick up all the saddle points otherwise.  We have relative minus sign
in the contributions from the non-trivial saddle points
$x_\ast =x_\pm$ because directions of the steepest descent through
$x_\ast =x_\pm$ are opposite between the cases for
$\pi/2 < {\rm arg}(g)<\pi$ and $-\pi < {\rm arg}(g)<-\pi /2$.  Note
that the real parts of the action at the non-trivial saddle points
change their signs across ${\rm arg}(g)=\pm3\pi/4$.  This means that
we have anti-Stokes lines of the small-$g$ expansion at
${\rm arg}(g)=\pm3\pi/4$.  This discussion is summarized in
fig.~\ref{fig:phi4_0d_stokes}. 
The large-$g$ expansion, on the other hand, is
independent\footnote{These behaviors of the expansions can be
  understood also from viewpoint of differential equation for the Bessel
  function although we can know this information after finding the
  exact result.  The modified Bessel function has essential
  singularity at $g=0$ and has only a branch cut singularity at
  $g=\infty$.  The Stokes phenomenon of the small-$g$ expansion is a
  manifestation of the essential singularity.  There is no such
  behavior near the branch point, and the large $g$ expansion is
  independent of ${\rm arg}(g)$.  } of ${\rm arg}(g)$:
\begin{equation}
  \label{eq:7}
  F(g)=g^{-1/2}\left( \frac{1}{2}  \Gamma ( 1/4) +\frac{1}{16} 
    \Gamma (-1/4 ) g^{-1} +\frac{1}{64}  \Gamma ( 1/4) g^{-2} 
    +\mathcal{O}(g^{-3}) \right) .
\end{equation}

\subsection{Interpolation along positive real axis}
\begin{figure}[t]
\begin{center}
\includegraphics[width=7.4cm]{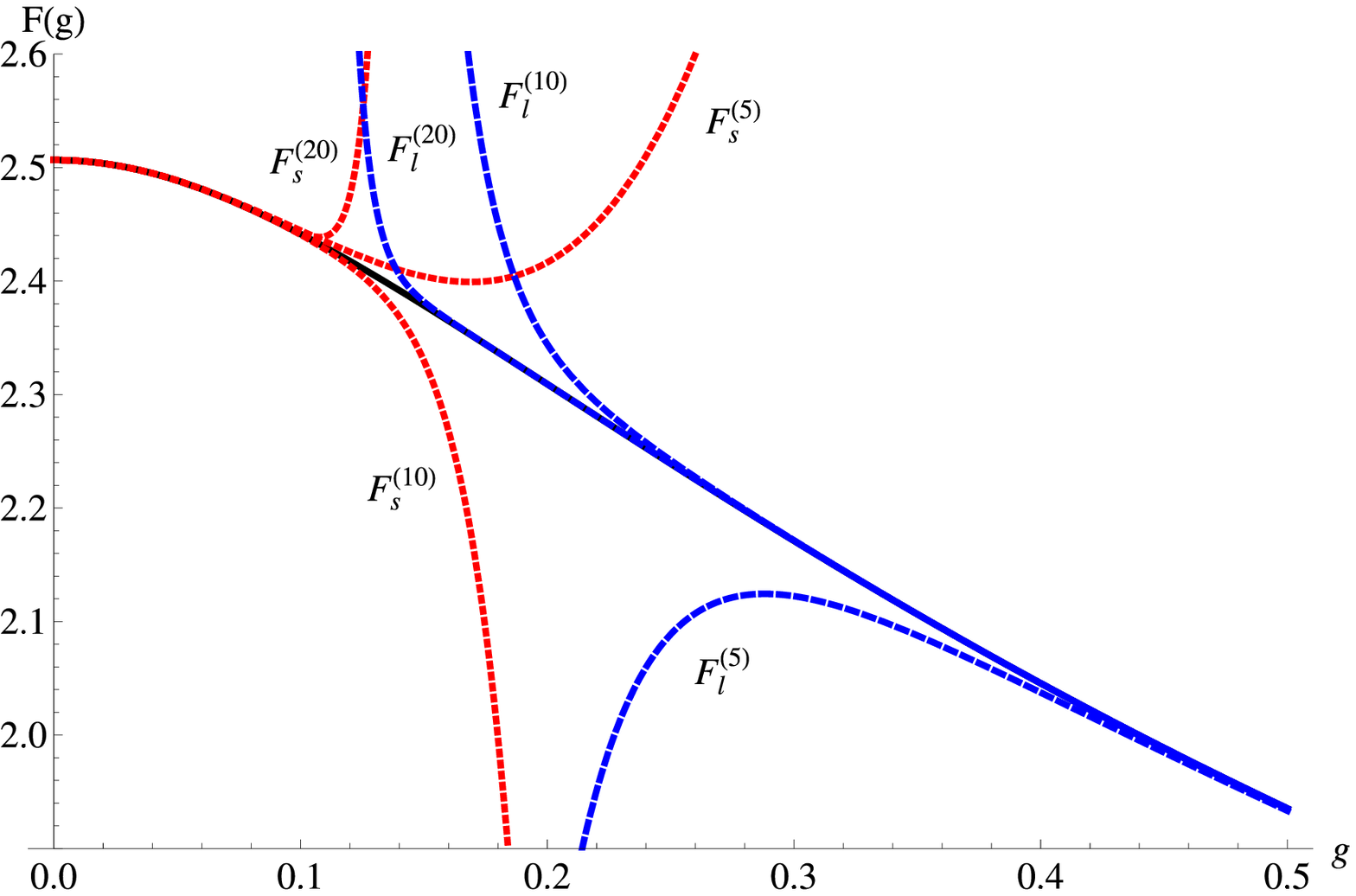}
\includegraphics[width=7.4cm]{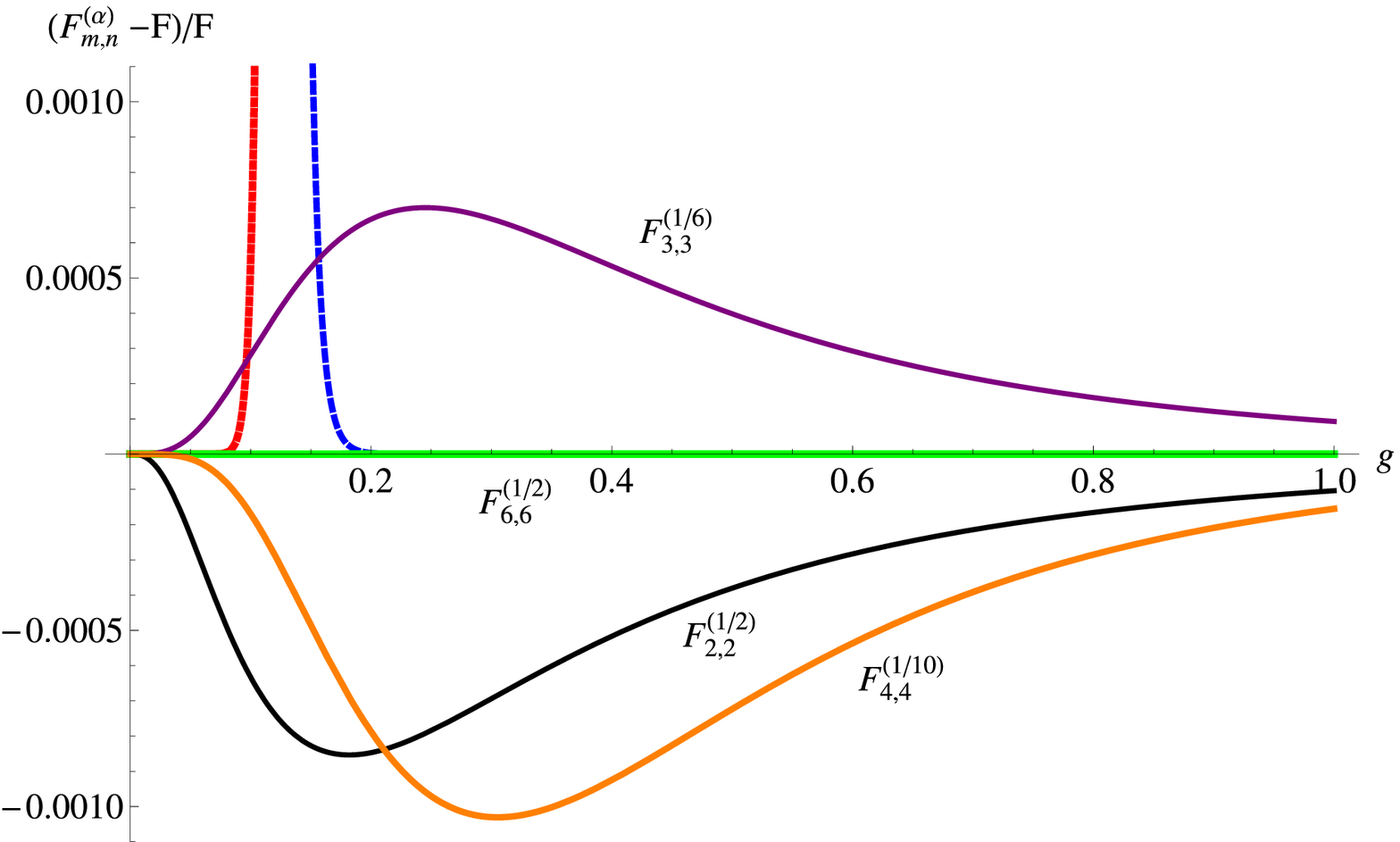}
\end{center}
\caption{
[Left] The partition function of the 0d $\varphi^4$ theory (black solid),
its small-$g$ expansions $F_s^{(N_s)}(g)$ (red dotted) and
large-$g$ expansions $F_l^{(N_l)}(g)$ (blue dashed) for ${\rm arg}(g)=0$.
[Right] Comparison of the FPRs with some $(m,n,\alpha )$ on the non-negative real axis of $g$.
}
\label{fig:phi4_0d_realp}
\end{figure}

Let us first take the coupling $g$ to be real and positive as usual.
For this case, we have the following small-$g$ and large-$g$
expansions
\begin{\eq}
 F_s^{(N_s )}(g) = \sum_{k=0}^{N_s} s_k g^k ,\quad s_{2k+1}=0 ,\quad s_{2k}=\frac{\sqrt{2}\Gamma (2k+1/2)}{k!} (-4)^k ,
\end{\eq}
\begin{\eq}
 F_{l}^{(N_l )}(g) = g^{-1/2}\sum_{l=0}^{N_l} l_k g^{-k} ,\quad 
l_k = \frac{\Gamma \left( \frac{k}{2} +\frac{1}{4} \right)}{2k!} \left( -\frac{1}{2} \right)^k ,
\end{\eq}
which are compared with the exact result in
fig.~\ref{fig:phi4_0d_realp} [Left].  In
terms of these expansions, we can construct FPR-type interpolating function
$F_{m,n}^{(\alpha )}(g)$ (see app.~\ref{app:0dphi4} for explicit
forms).  In fig.~\ref{fig:phi4_0d_realp} [Right], we test validity of
the FPRs by plotting
\[
\frac{F_{m,n}^{(\alpha )}-F(g)}{F(g)} ,
\]
against $g$ for some $(m,n,\alpha )$.  We easily see that these
interpolating functions provide good approximations to the original
function $F(g)$.  Especially $F_{6,6}^{(1/2)}(g)$ approximates the
exact result $F(g)$ very well: the maximal value of the ratio
$|(F_{6,6}^{(1/2 )}(g)-F(g))/F(g)|$ is $\mathcal{O}(10^{-8})$.  Thus
we find that our interpolating scheme in this example works quite well
at least along the positive real axis of $g$.  Of course this result
is not new and has been already seen in the previous studies
\cite{Sen:2013oza,Honda:2014bza}.  Here we ask another question.
Suppose we perform naive analytic continuation of the interpolating
function $F_{m,n}^{(\alpha )}(g)$ to the whole complex plane of $g$.
Then, {\it does the interpolating function $F_{6,6}^{(1/2)}(g)$, which
is very precise along the positive real $g$, still gives a good
approximation beyond the positive real axis?}

\begin{figure}[t]
\begin{center}
\includegraphics[width=7.0cm]{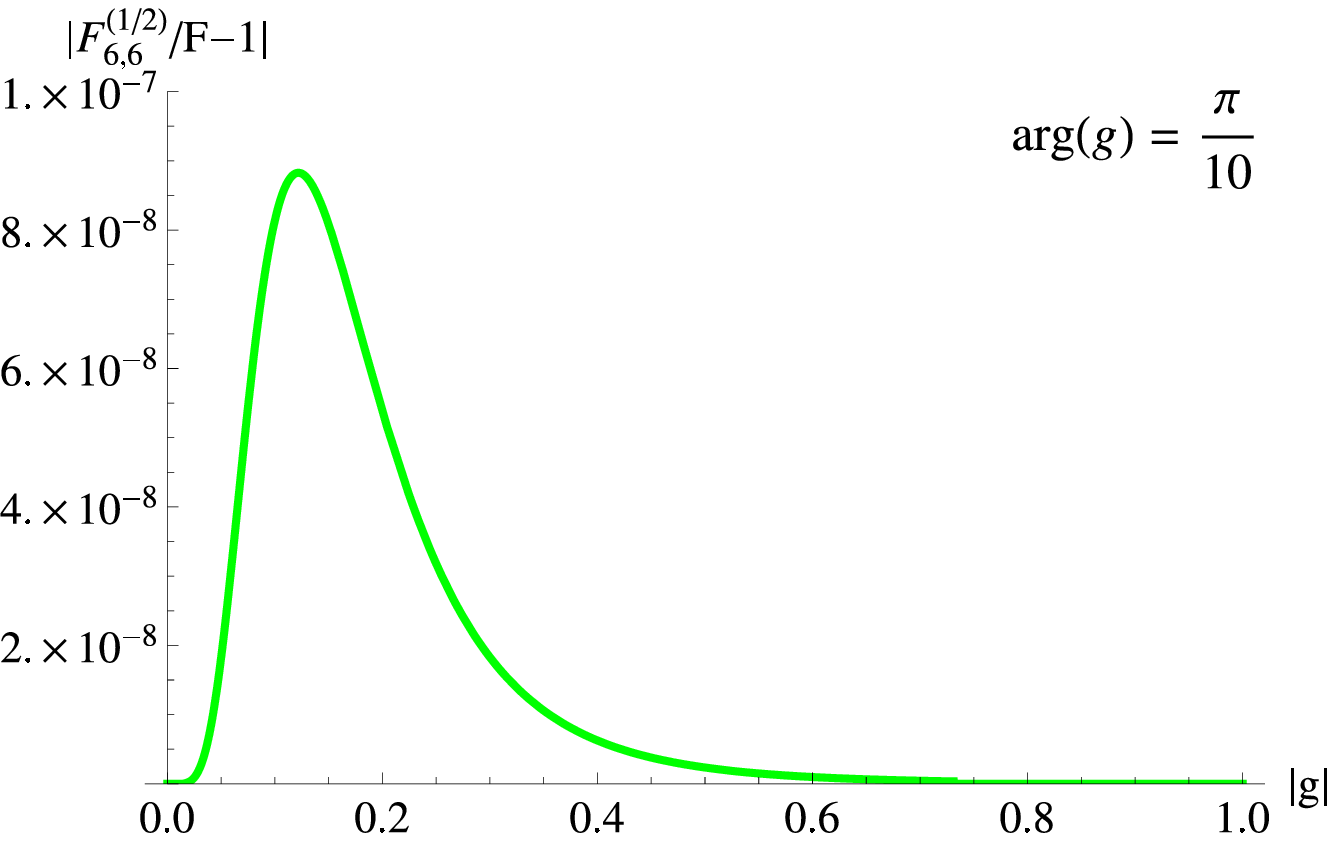}
\includegraphics[width=7.0cm]{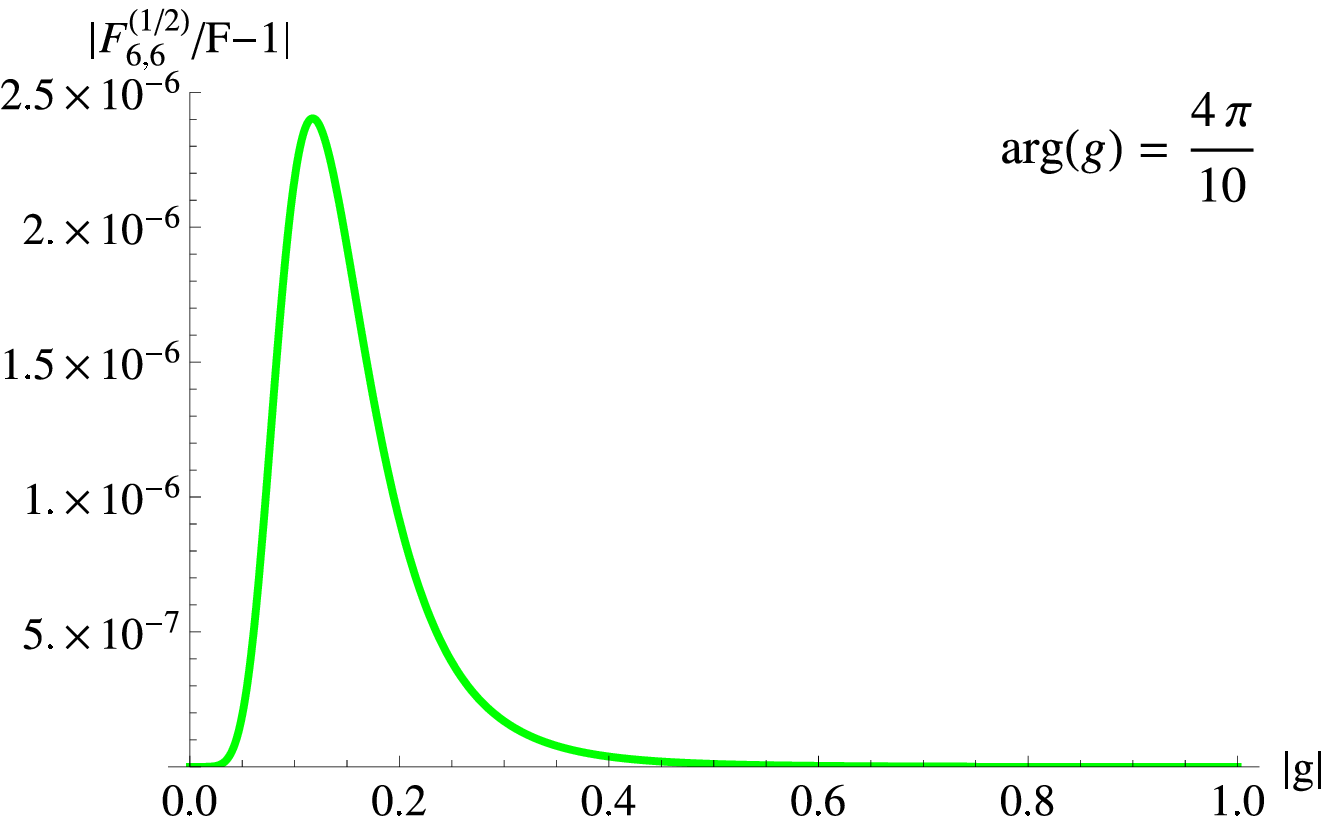}\\
\includegraphics[width=7.0cm]{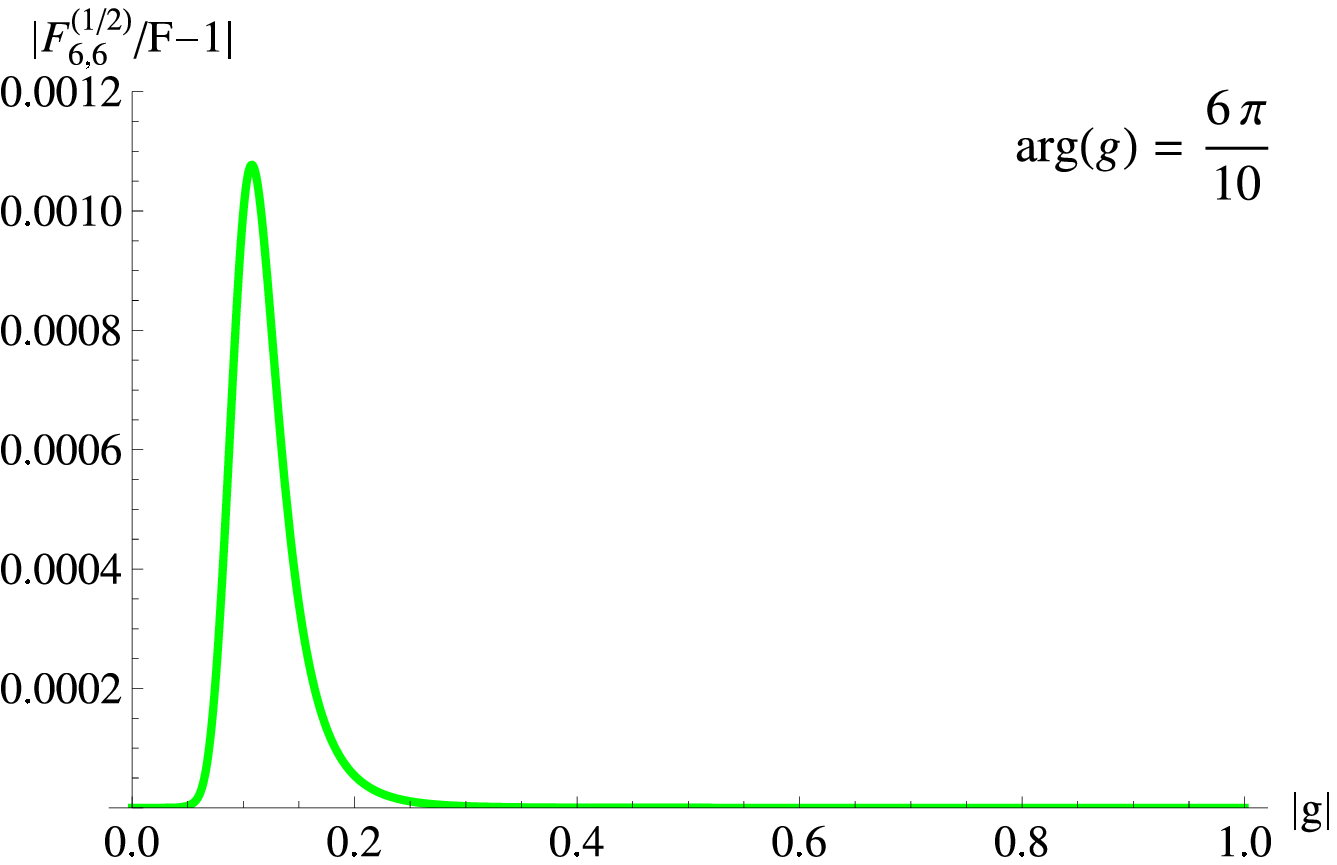}
\includegraphics[width=7.0cm]{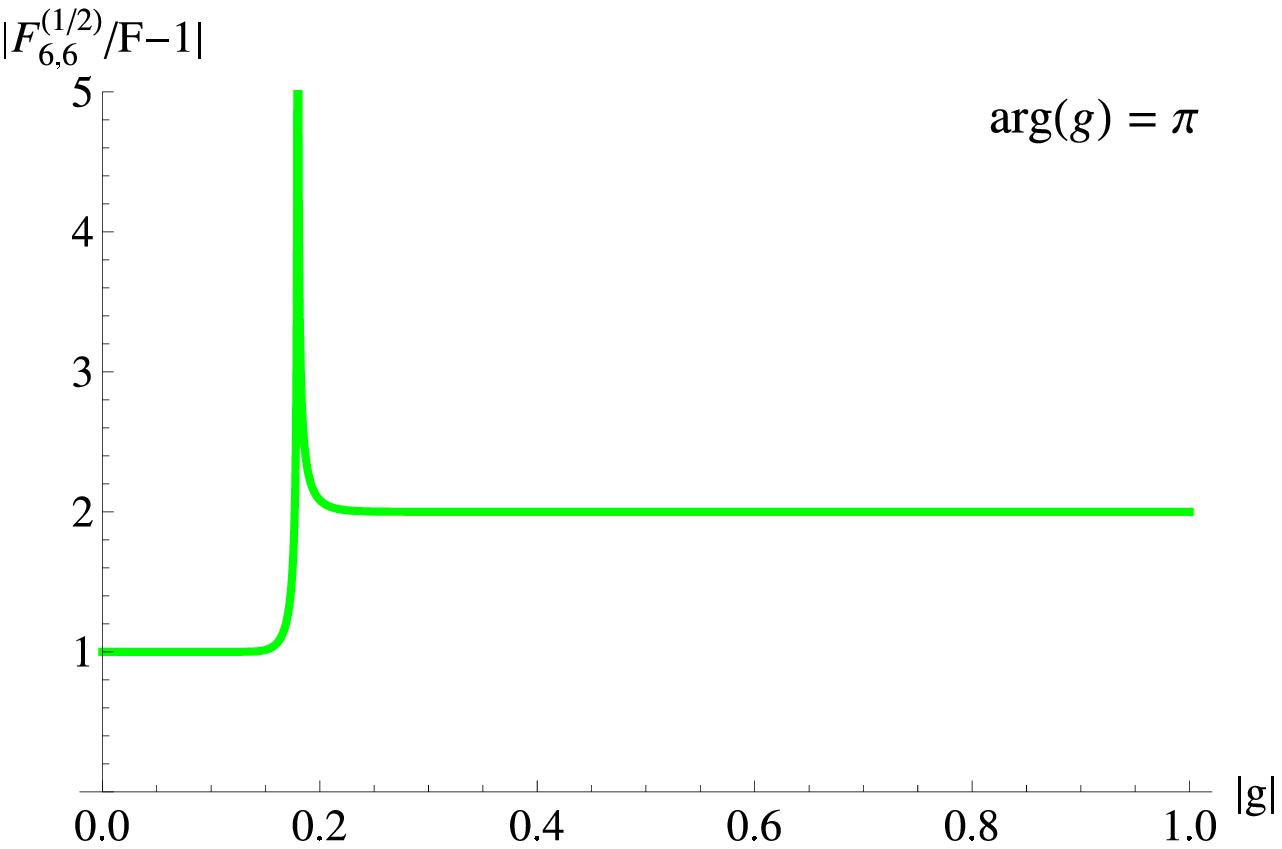}
\end{center}
\caption{The quantity $|(F_{6,6}^{(1/2 )}(g)-F(g))/F(g)|$ is plotted
  against $|g|$ for some ${\rm arg}(g)$.}
\label{fig:phi4_0d_rp_arg}
\end{figure}

\begin{figure}[tbp]
\begin{center}
\includegraphics[width=7.4cm]{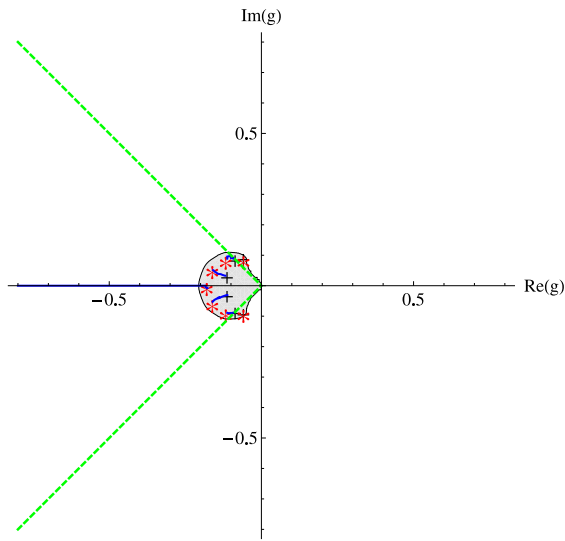}
\end{center}
\caption{The region where the interpolating function
  $F_{6,6}^{(1/2)}(g)$ gives bad approximation.  In the shaded region,
  the ratio $|F_{6,6}^{(1/2)}/F-1|$ is larger than $0.05$.  We also
  plot ``zeros'' (the symbol ``$+$") and ``poles'' (the symbol
  ``$\ast$") of the rational function $(F_{6,6}^{(1/2)}(g))^2$
  associated with the FPR.  The blue solid lines denote the branch
  cuts of $F_{6,6}^{(1/2)}(g)$.  }
\label{fig:phi4_0d_rp_complex}
\end{figure}

In order to answer this question, we plot the quantity\footnote{Note
  that $(F_{6,6}^{(1/2 )}(g)-F(g))/F(g)$ does not take real value for
  complex $g$ in general.  }  $|F_{6,6}^{(1/2 )}(g)/F(g)-1|$ against
$|g|$ for some ${\rm arg}(g)$ in fig.~\ref{fig:phi4_0d_rp_arg}.  
First, we easily observe in
fig.~\ref{fig:phi4_0d_rp_arg} [Left top] that the FPR gives very
precise approximation for ${\rm arg}(g)=\pi /10$, whose relative error
is $\mathcal{O}(10^{-7})$ at worst.  Fig.~\ref{fig:phi4_0d_rp_arg}
[Right top] shows that this is true also for ${\rm arg}(g)=4\pi /10$,
albeit not as accurate as that for ${\rm arg}(g)=\pi /10$.  We
therefore conclude that the FPR can give good approximation
of the exact function even beyond the real positive axis.  
However, as we further increase ${\rm arg}(g)$ to $\pi$, the approximation starts
becoming worse.  
As seen in fig.~\ref{fig:phi4_0d_rp_arg} [Left
bottom], the FPR still gives good approximation for
${\rm arg}(g)=6\pi /10$ but the relative error becomes $\mathcal{O}(0.1\% )$.  
Since we have the exponentially suppressed corrections in the weak coupling regime 
for $\pi/2 < {\rm arg}(g)<3\pi /4$, which comes from the nontrivial saddle
points, we recognize that the exponentially suppressed corrections are
responsible for this error.  For $3\pi/4 < {\rm arg}(g)<5\pi /4$, the
contributions from the nontrivial saddle points become exponentially growing.  
Since the FPR lacks this information, the FPR should show
very large error for $3\pi/4 < {\rm arg}(g)<5\pi /4$ in small-$|g|$ regime.  
Indeed we have $\mathcal{O}(100\% )$ error on the negative real axis as seen in
fig.~\ref{fig:phi4_0d_rp_arg} [Right bottom].  
In fig.~\ref{fig:phi4_0d_rp_complex}, we summarize validity of approximation by the FPR.  
The shaded part shows the region where the FPR has more
than $5\%$ relative error.  We also draw the zeros and poles of the
rational function $(F_{6,6}^{(\alpha )})^2$ associated\footnote{ If we
  consider general FPR $F_{m,n}^{(\alpha )}$, then its natural
  associated rational function is
  $(F_{m,n}^{(\alpha )})^{1/|\alpha |}$.  } with the FPR, which give
branch cuts of the FPR.
From this figure we observe some points:
\begin{itemize}
\item The shaded region looks like a fan, whose radial bounding lines
  are close to the anti-Stokes lines of the small-$g$ expansion.  This
  is natural since the dominant part of the small-$g$ expansion
  changes across the anti-Stokes lines and the FPRs do not know this
  information in small-$|g|$ regime.  The radius of the fan should be
  finite since the FPR gives the correct large-$g$ expansion by
  construction even across the anti-Stokes lines.

\item The boundary of the (shaded) fan-like region is similar to the
  region surrounded by the origin, poles and zeros of the rational function $(F_{6,6}^{(1/2)}(g))^2$. 
  Especially the anti-Stokes lines are close to the lines between the origin and the
  first poles measured from the positive real axis\footnote{The first poles
  are located at ${\rm arg}(g)\simeq \pm2.358$ while the anti-Stokes
  lines are oriented along ${\rm arg}(g)=3\pi /4\simeq 2.356$.}.  
  This would be natural because when the exact function $F(g)$ does not
  have singularities around the poles of $(F_{6,6}^{(1/2)}(g))^2$, then the FPR differs from
  the exact function by a large amount in the neighborhood of its
  poles.  Hence for this case, the FPR clearly gives bad approximation
  around the poles.

\item There is a branch cut from the pole at the negative real axis to
  $g=-\infty$.  Since the partition function $F(g)$ has the branch cut
  on $(0,-\infty )$, we interpret that the branch cut of the FPR
  approximates the ``true" branch cut of the exact result.

\item Although the small-$g$ expansion has the Stokes line at the
  imaginary axis, the FPR does not detect this Stokes line.  This is
  because the dominant part of the small-$g$ expansion does not change
  across this Stokes line and only the sub-dominant part changes.

\end{itemize}
These results lead us to the conjecture about the general feature of
FPR that each branch cut of the FPR has the following possible interpretations.
\begin{enumerate}
\item The branch cut is an artifact of the approximation by the FPR.

\item The physical quantity $F(g)$ has a branch cut near the branch cut of the FPR.

\item Near the branch cut, one of perturbation series of $F(g)$
  changes its dominant part.  This case implies the following two possibilities
  on Stokes phenomena.
\begin{enumerate}
\item We have an {\it anti-Stokes line} of the perturbative expansion
  near the branch cut.  This possibility occurs most likely for the
  first branch cut measured from the specific axis where the interpolating function is constructed.

\item We have a {\it Stokes line} near the branch cut, whose diagonal
  multiplier is different from 1.
\end{enumerate}
\end{enumerate}

One of immediate questions here is if these features are particular
for this problem or true also for other problems.  We will explicitly
check this in sec.~\ref{sec:SG} that this is true also for the partition functions
of the 0d Sine-Gordon model.  This seems to hold also in other
examples such as BPS Wilson loop in 4d $\mathcal{N}=4$ Super
Yang-Mills theory, energy spectrum in 1d anharmonic oscillator etc
\cite{progress}.  Another important question is if we can construct
another interpolating function, which gives good approximation in the
region, where the interpolating function along the positive real axis
becomes bad.  One natural way to do this is to construct interpolating
functions along a specific axis with $|{\rm arg}(g)|>3\pi /4$, where
dominant part of the small-$g$ expansion comes from the non-trivial
saddle points, and then to extend these to the complex $g$.  In next
subsection, we will perform this by considering interpolating functions
along the {\it negative} real axis.

\subsubsection*{Remarks}
\begin{figure}[tbp]
\begin{center}
\includegraphics[width=7.4cm]{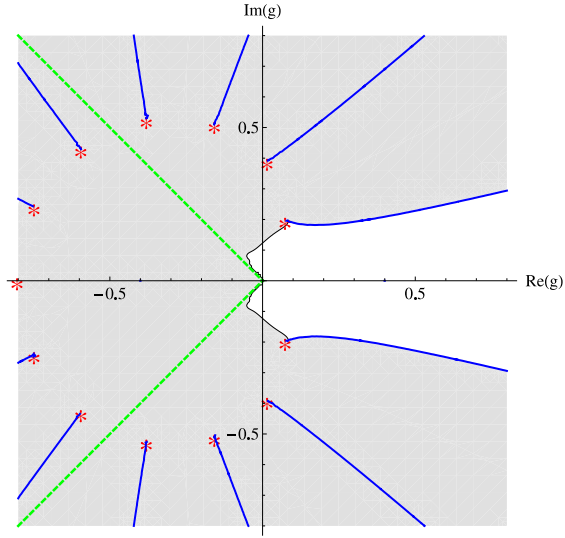}
\end{center}
\caption{Similar plot as fig. \ref{fig:phi4_0d_rp_complex} for the FPP $F_{6,6}^{(1/26)}(g)$.}
\label{fig:phi4_0d_rp_complex_FPP}
\end{figure}

One might ask if the FPRs $F_{m,n}^{(\alpha )}$
with the same $(m,n)$ but different $\alpha$ gave similar results.  In
this example we find that the results strongly depend on $\alpha$.
For instance, let us look at fig.~\ref{fig:phi4_0d_rp_complex_FPP},
which is similar to the plot in fig.~\ref{fig:phi4_0d_rp_complex} but
for $F_{6,6}^{(1/26)}(g)$.  Note that the FPR for this case becomes
the FPP \cite{Sen:2013oza}.  On the real positive axis, this
interpolating function gives maximum error of about $0.6\%$.  It is
easy to see from fig.~\ref{fig:phi4_0d_rp_complex_FPP} that the
analytic structure of $F_{6,6}^{(1/26)}(g)$ is very different from the
one of $F_{6,6}^{(1/2)}(g)$.  Especially, when we start from the real
positive axis and go towards the anti-Stokes line at ${\rm arg}(g)=3\pi /4$, we
encounter many branch cuts for this case.  Another important
difference is that the first branch cuts from the real positive axis are
located at the pretty smaller angle ${\rm arg}(g)\simeq \pm 1.21$, which
is quite far from the location of the anti-Stokes likes.  This
seems to be the reason why the FPP $F_{6,6}^{(1/26)}(g)$ gives the
worse approximation than $F_{6,6}^{(1/2)}(g)$.

\subsection{Interpolation along negative real axis}
\label{sec:phi4_negative}
\begin{figure}[t]
\begin{center}
\includegraphics[width=7.4cm]{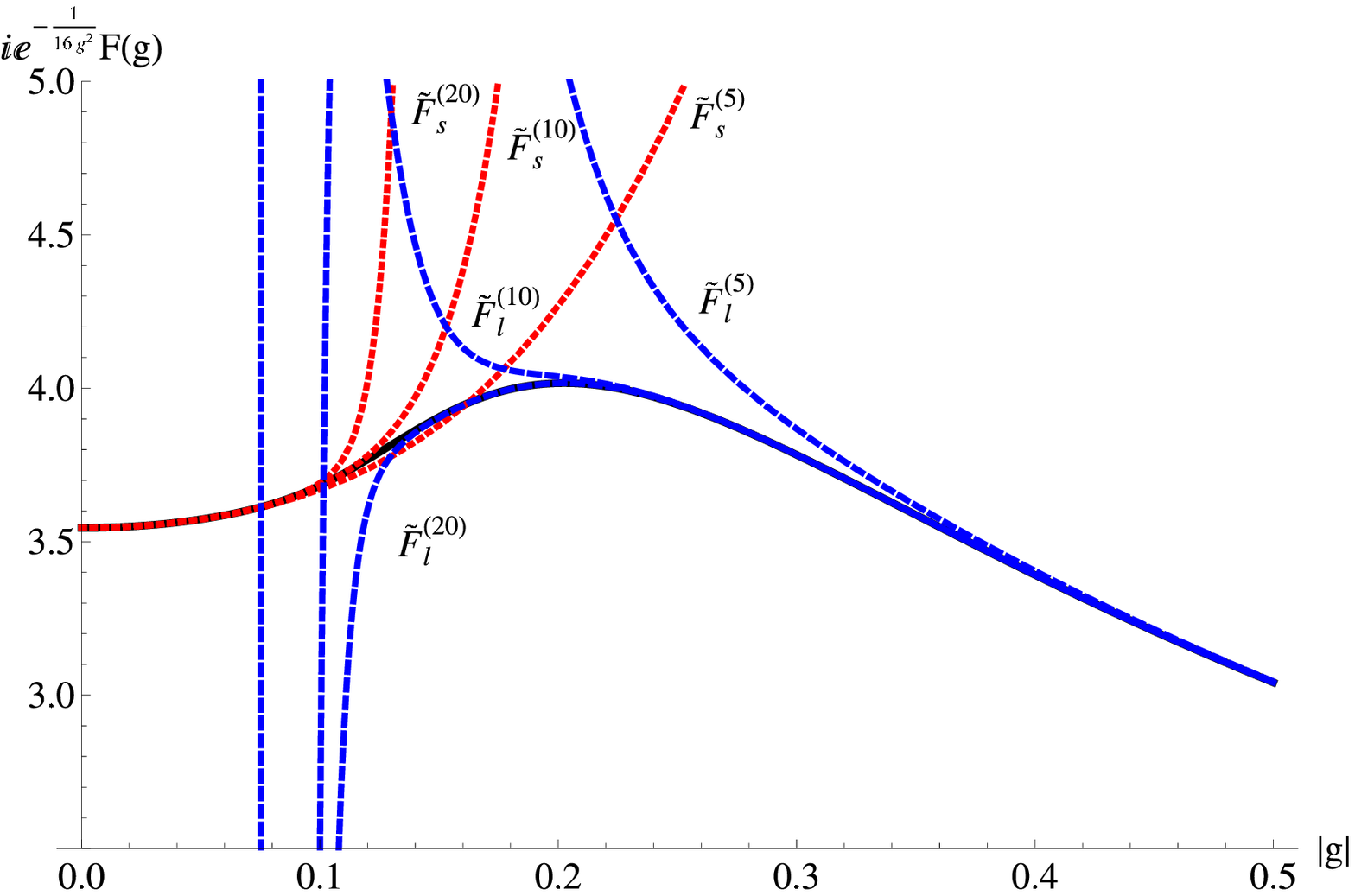}
\includegraphics[width=7.4cm]{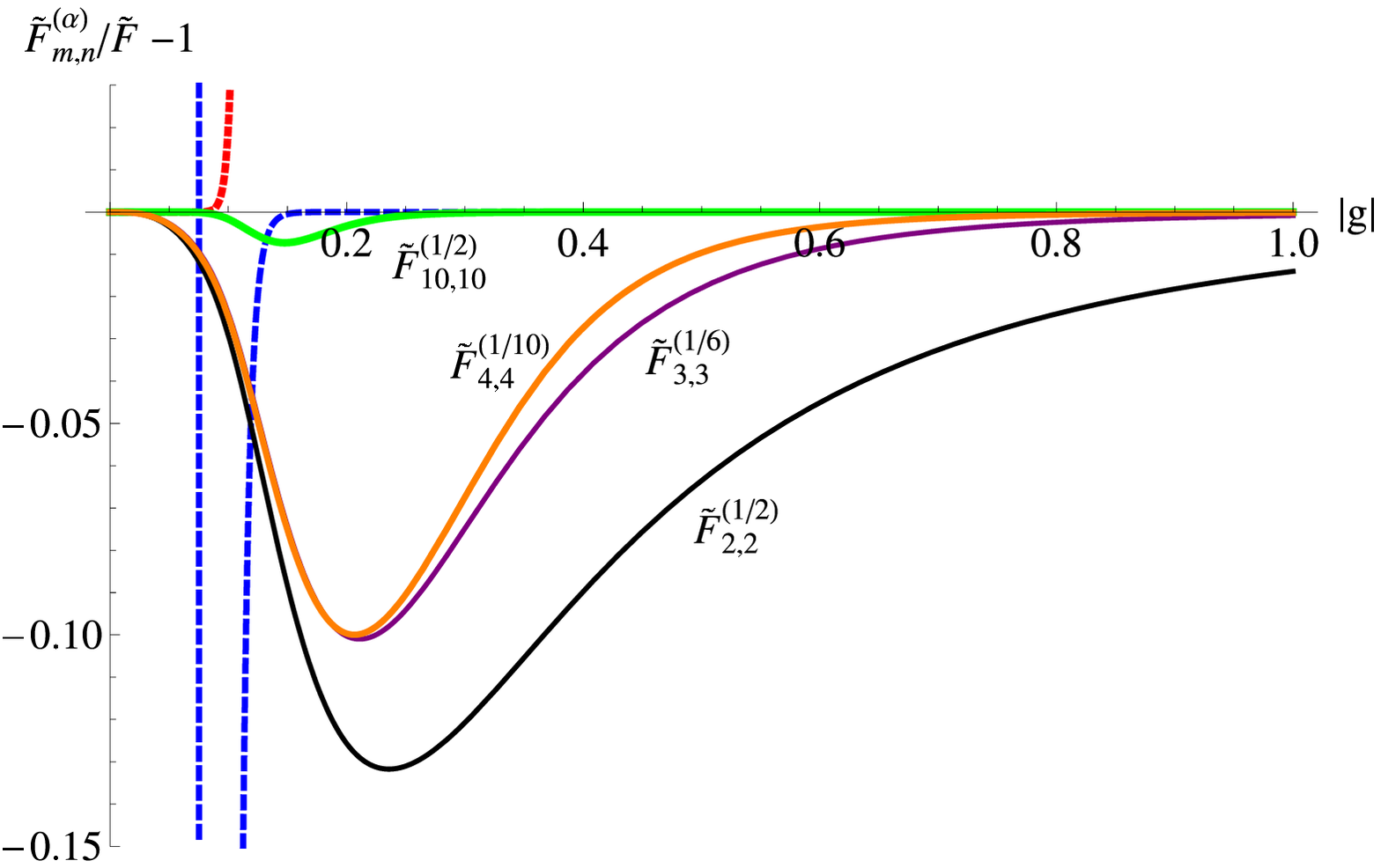}
\end{center}
\caption{ [Left] The function $\tilde{F}(g)=ie^{-\frac{1}{16g^2}}F(g)$
  (black solid), its small-$g$ expansions $\tilde{F}_s^{(N_s)}(g)$
  (red dotted) and large-$g$ expansions $\tilde{F}_l^{(N_l)}(g)$ (blue
  dashed) for ${\rm arg}(g)=\pi -\epsilon$.  [Right] The quantity
  $(\tilde{F}_{m,n}^{(\alpha )}(t)/\tilde{F}(t)-1)
  =(F_{L+,m,n}^{(\alpha )}(-t)/F(-t)-1)$ is plotted to $t=|g|$.}
\label{fig:phi4_0d_realn}
\end{figure}
In order to construct another interpolating function precise for
$|{\rm arg}(g)|>3\pi /4$, let us consider interpolating function along
the negative real axis of $g$ and then consider its naive analytic
continuation to the whole complex plane.  The function $F(g)$ has the
following small-$g$ and large-$g$ expansions on
$\epsilon$-neighborhood of the negative real axis:
\begin{\eqa}
F(-t+i\epsilon )
&=&-\sqrt{\pi } i e^{\frac{1}{16 t^2}} \left( 2 +6 t^2+\mathcal{O}(t^4 ) \right)
+\sqrt{\frac{\pi}{2} } \left( 2 -6 t^2  +\mathcal{O}(t^4 ) \right) +\mathcal{O}(\epsilon ) \NN\\
&=& -i t^{-1/2}
 \left( \frac{1}{2}  \Gamma ( 1/4) -\frac{1}{16} \Gamma (-1/4 ) t^{-1} +\frac{1}{64}  \Gamma ( 1/4) t^{-2} +\mathcal{O}(t^{-3}) \right) 
+\mathcal{O}(\epsilon ) , \NN\\
F(-t-i\epsilon )
&=&+\sqrt{\pi } i e^{\frac{1}{16 t^2}} \left( 2 +6 t^2+\mathcal{O}(t^4 ) \right) 
+\sqrt{\frac{\pi}{2} } \left( 2-6 t^2  +\mathcal{O}(t^4 ) \right) +\mathcal{O}(\epsilon ) ,\NN\\
&=& +i t^{-1/2}
 \left( \frac{1}{2}  \Gamma ( 1/4) -\frac{1}{16} \Gamma (-1/4 ) t^{-1} +\frac{1}{64}  \Gamma ( 1/4) t^{-2} +\mathcal{O}(t^{-3}) \right) 
+\mathcal{O}(\epsilon ) , \NN\\
\end{\eqa}
with $t\in\mathbb{R}_+$.  Namely, the dominant parts of the small-$g$
expansion and the large-$g$ expansion change their signs across the
negative real axis.  This reflects that the exact function $F(g)$ has
the square root branch cut on the negative real axis.  Instead of
$F(g)$, let us consider interpolating functions of the quantity:
\begin{\eq}
\tilde{F}(t) =\lim_{\epsilon\rightarrow +0}
\left.  i e^{-\frac{1}{16g^2}} F(g+i\epsilon ) \right|_{g\rightarrow -t}, \quad t\in\mathbb{R}_+ .
\end{\eq}
The function $\tilde{F}(t)$ has the small-$t$ and large-$t$ expansions,
\begin{\eqa}
\tilde{F}(t)
&=& 2 \sqrt{\pi }+6 \sqrt{\pi } t^2+105 \sqrt{\pi } t^4+3465 \sqrt{\pi } t^6+\frac{675675 \sqrt{\pi } t^8}{4} +O\left(t^{10}\right) \NN\\
&=& t^{-1/2} \left( 
\frac{\pi  }{\sqrt{2} \Gamma \left(\frac{3}{4}\right)}  +\frac{\pi t^{-1}}{8 \sqrt{2} \Gamma \left(\frac{5}{4}\right)} 
-\frac{\pi t^{-2}}{32 \sqrt{2} \Gamma \left(\frac{3}{4}\right)} 
-\frac{\pi t^{-3}}{256\sqrt{2} \Gamma \left( \frac{5}{4} \right) }
 +O\left( t^{-4}\right) \right) ,
\end{\eqa}
where we have dropped the exponentially suppressed correction
$\mathcal{O}(e^{-\frac{1}{16t^2}})$ coming from the trivial saddle
point in the small-$t$ expansion.  
Denoting interpolating function of $\tilde{F}(t)$ as $\tilde{F}_{m,n}^{(\alpha )}(t)$, 
we can approximate the original function by using the function
\begin{\eq}
F_{L\pm ,m,n}^{(\alpha )} (g) =  \mp i e^{+\frac{1}{16g^2}} \tilde{F}_{m,n}^{(\alpha )} (-g) .
\end{\eq}
The interpolating function $F_{L\pm ,m,n}^{(\alpha )} (g) $ reproduces
the small-$g$ and large-$g$ expansions of $F(g)$ for
$g\in \mathbb{R}_- \pm i\epsilon$ up to certain orders.  Indeed
$F_{L+ ,m,n}^{(\alpha )} (g) $ approximates $F(g)$ quite well along
the negative real ($+i\epsilon $) axis\footnote{
Similar result holds also for $F_{L- ,m,n}^{(\alpha )} (g) $.
} as seen in fig.~\ref{fig:phi4_0d_realn}
[Right].  Especially $F_{L\pm ,10,10}^{(\alpha )} (g) $ has about
relative $0.8\%$ error at worst.

\begin{figure}[t]
\begin{center}
\includegraphics[width=7.4cm]{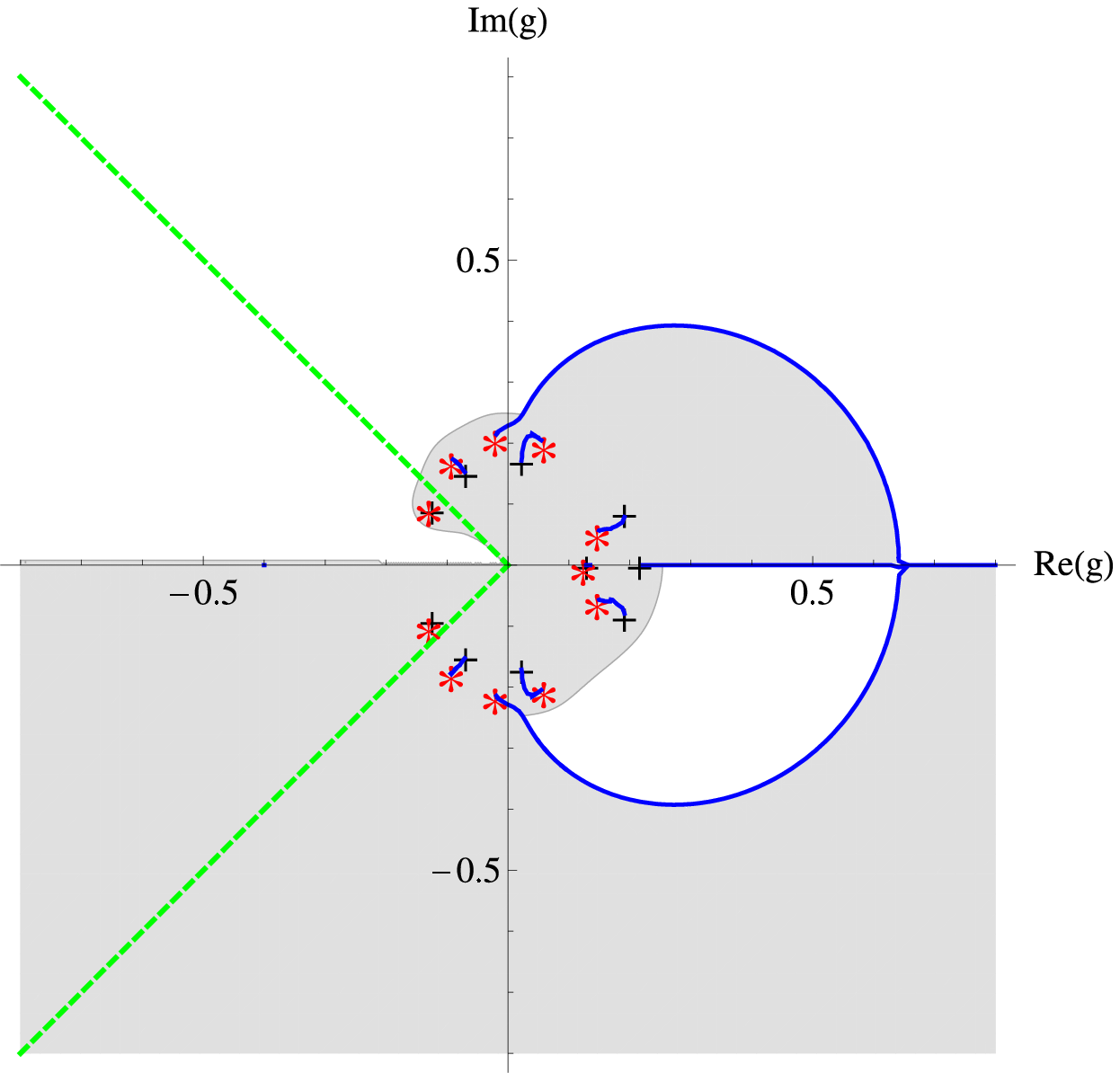}
\includegraphics[width=7.4cm]{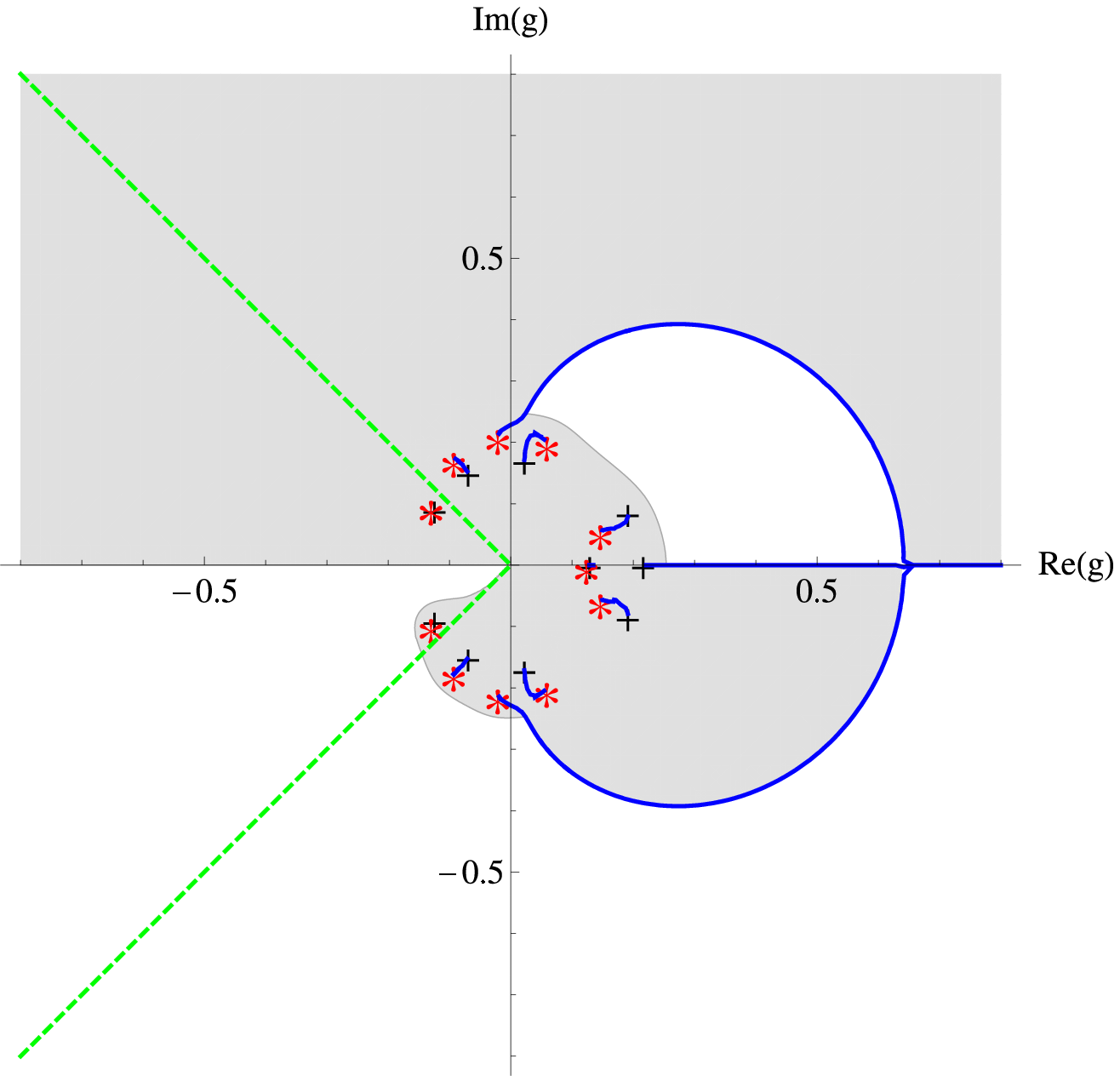}
\end{center}
\caption{[Left] The region where the ratio
  $|F_{L+,10,10}^{(1/2)}/F-1|$ is larger than $0.05$.  
  We also plot zeros ($+$) and poles ($\ast$)
  of the rational function $(\tilde{F}_{10,10}^{(1/2)})^2$.  The blue
  solid lines denote the branch cuts of $F_{L+,10,10}^{(1/2)}$.
  [Right] Similar plot as the left for $F_{L-,10,10}^{(1/2 )}$.  }
\label{fig:phi4_0d_rn_complex}
\end{figure}

Let us consider general complex $g$ in the interpolating function
$F_{L\pm ,m,n}^{(\alpha )} (g) $ as in the last subsection.  Then,
unless we cross the anti-Stokes line or branch cuts,
$F_{L\pm ,m,n}^{(\alpha )} (g) $ gives the correct large-$g$ expansion
and dominant part of small-$g$ expansion.  If we cross the anti-Stokes
line, then the small-$g$ expansion is dominated by the contribution
from the trivial saddle point and $F_{L\pm ,m,n}^{(\alpha )} (g) $
should fail to approximate $F(g)$ in small-$|g|$ regime.  
Also, if we cross the branch cut particular to the FPR, 
then the FPR will pick up an extra phase and also break the approximation.
This extra phase could be trivial depending on the
number of times, where we cross branch cuts.

The validity of the approximation by $F_{L\pm ,10,10,}^{(-1/2 )} (g)$
is summarized with its analytic property in fig.~\ref{fig:phi4_0d_rn_complex}.
One of important differences from the interpolation along
the real positive axis is that the (semi-)circular branch cut of
$F_{L\pm ,10,10}^{(1/2 )} (g)$ surrounds all the poles and zeros on
the right plane.  Therefore if we go across the circular branch cut on
the right half plane, then $F_{L\pm ,10,10}^{(1/2 )} (g)$ undergoes
a sign flip and hence it fails to approximate $F(g)$ across the
circular branch cut.  However, these FPRs also have a branch cut along
the positive real axis and crossing this branch cut on the positive
real axis leads to another flip in the sign.  As a result
$F_{L\pm ,10,10}^{(1/2 )} (g)$ recovers the correct sign and gives 
the good approximation to $F(g)$ again.  That is why we see the
disconnected unshaded regions in fig.~\ref{fig:phi4_0d_rn_complex}.

\begin{figure}[t]
\begin{center}
\includegraphics[width=8.0cm]{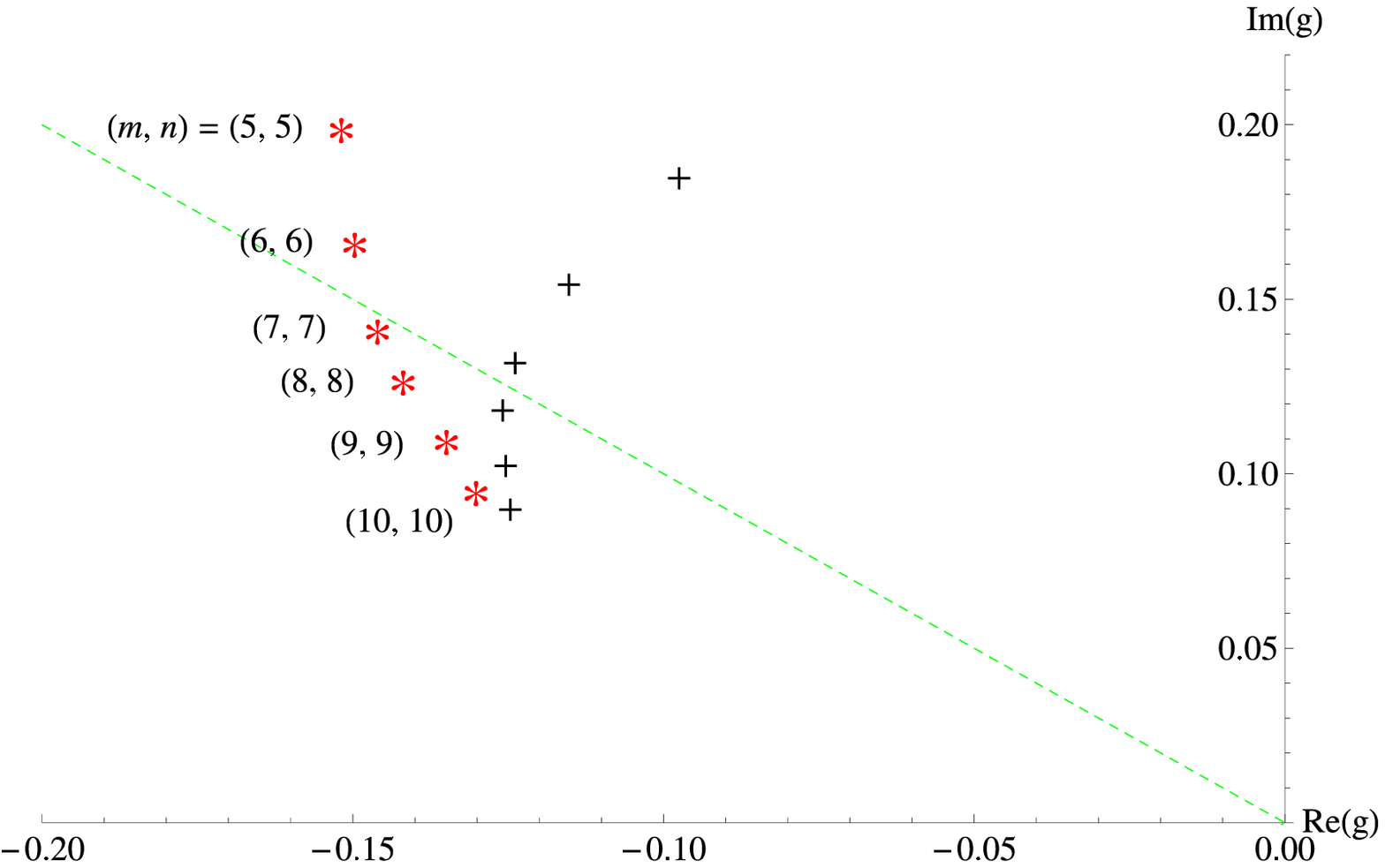}
\includegraphics[width=7.4cm]{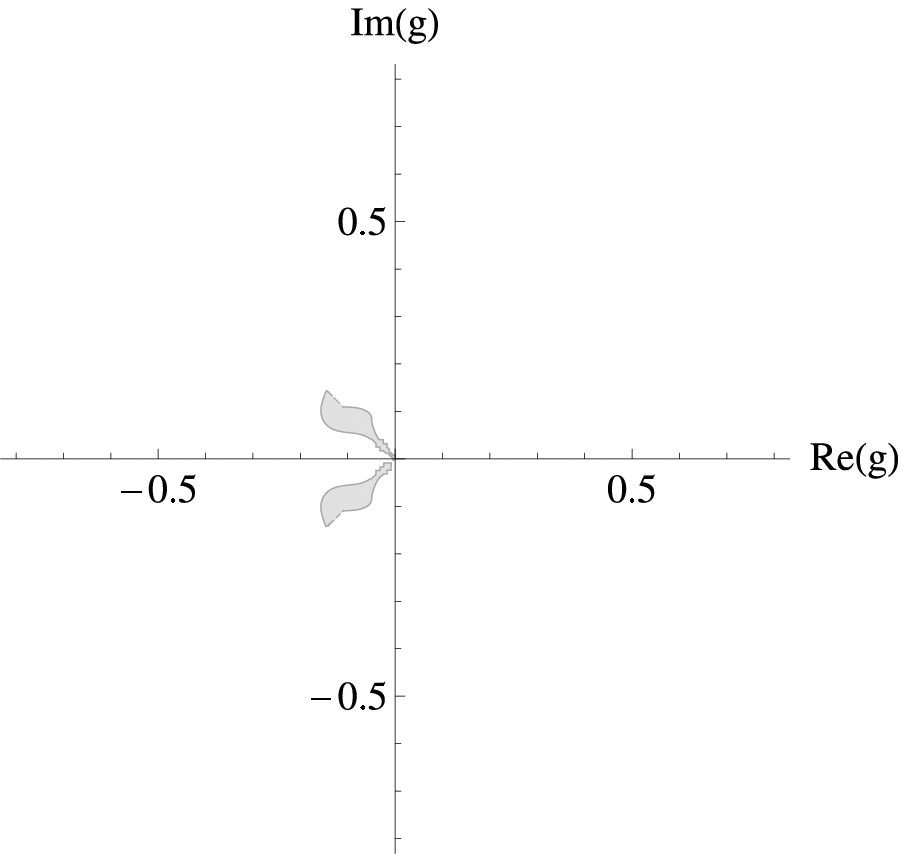}
\end{center}
\caption{[Left] First poles and zeros of $(F_{m,n}^{(1/2 )})^2$ from
  the negative real axis for various $(m,n)$.  The symbol ``+''
  denotes zeros and ``*'' denotes poles.  [Right] The region where the
  patch $F'(g)$ of the best FPRs gives more than relative $5\%$
  error.}
\label{fig:phi4_0d_1st_poles}
\end{figure}

Another important difference is that the first branch cuts as measured
from the negative real axis slightly deviate from the anti-Stokes lines.  
While the anti-stokes lines are oriented
along ${\rm arg}(g)= \pm3\pi /4 \simeq \pm 2.356$, these first cuts
are located at the angle\footnote{Note that 
when we write values of ${\rm arg}(g)$, we always denote the values measured with respect to the positive
real axis with counterclockwise. } ${\rm arg}(g)=\pm2.504$.  However, we expect
that as increasing the values of $(m,n)$ in the FPR $F_{m,n}^{(1/2)}$,
the first poles of $(F_{m,n}^{(1/2)})^2$ will approach to the first zeros
from the negative real axis and the first cuts will finally vanish for
sufficiently large $(m,n)$.  Indeed we can easily observe in
fig.~\ref{fig:phi4_0d_1st_poles} [Left] that the pair of first zeros and
poles seem to converge to the same point as increasing $(m,n)$.  
Thus we conclude that the first pair of the branch cuts are the artifact of the
approximation by the FPR with insufficiently large $(m,n)$.  We call
this type of singularities ``fake singularities".  The above result
would be natural because the FPR with larger $(m,n)$ tends to give
better approximation in this problem and may improve the validity of
the approximation near the anti-Stokes lines\footnote{ When either of
  the small-$g$ or large-$g$ expansion is convergent as in this
  problem, we expect this tendency because the convergent expansion
  itself gives very precise approximation inside its radius of
  convergence and we can regard their FPRs with large $(m,n)$ as
  analytic continuation of the convergent expansion to whole range of
  values of $g$.  However, it is nontrivial in general whether the
  decreasing behavior of errors in the FPRs will be monotonic or not,
  although the error of the particular FPR $F_{m,n}^{(1/2)}$ in this
  problem seems to be monotonically decreasing.  For example, in 2d
  Ising model with finite volume, both low and high temperature
  expansions of specific heat are convergent For this case, error of
  its FPR with fixed $\alpha$ tends to decrease but not monotonically
  decrease as increasing $(m,n)$ \cite{Honda:2014bza}.  }.  In next
section we will see that FPRs in the 0d Sine-Gordon model have similar
features.

Finally let us find good approximation of $F(g)$ in region as wide as
possible by patching the best interpolating functions along the
positive and negative real axis.  In fig,~\ref{fig:phi4_0d_1st_poles}
[Right], we draw range of validity of approximation by
\begin{\eq}
F' (g) = \left\{ \begin{matrix}
F_{6,6}^{(1/2)} (g)        & {\rm for}& |{\rm arg}(g)| <3\pi /4 \cr
F_{L+,10,10}^{(1/2)} (g) & {\rm for}& 3\pi /4 < {\rm arg}(g) <\pi \cr
F_{L-,10,10}^{(1/2)} (g) & {\rm for}& -\pi  < {\rm arg}(g) < -3\pi /4 \cr
\end{matrix}\right. .
\end{\eq}
This indicates that the patching $F' (g)$ has $5\%$ or better accuracy
in the very wide region.

\section{Partition function of zero-dimensional Sine-Gordon
  model}\label{sec:SG}
\begin{figure}[tbp]
\begin{center}
\includegraphics[width=7.4cm]{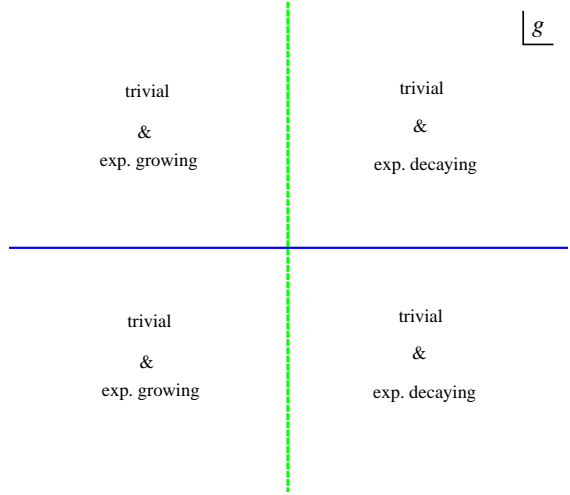}
\end{center}
\caption{Summary of Stokes phenomena for the small-$g$ expansion 
in the partition function of the 0d Sine-Gordon model.
The blue solid lines denote the Stokes lines
while the green dashed lines denote the anti-Stokes lines.}
\label{fig:toymodel_0d_stokes}
\end{figure}
Let us consider the partition function of the zero-dimensional Sine-Gordon
model:
\begin{\eq}
F(g) =\frac{1}{\sqrt{g}}\int_{-\pi /2}^{\pi /2} dx\ 
e^{-\frac{1}{2g}\sin^2{x}} ,
\end{\eq}
which was considered by Cherman-Koroteev-Unsal in the context of
resurgence \cite{Cherman:2014xia}.  As in the last section, this integral
can be evaluated exactly as
\begin{\eq}
F(g) =\frac{\pi}{\sqrt{g}} e^{-\frac{1}{4g}} I_0 \left( \frac{1}{4g} \right) .
\end{\eq}
The function $F(g)$ has the following small-$g$ and large-$g$ expansions
\begin{\eq}
F(g)
= \left\{ \begin{matrix}
  \sqrt{\frac{\pi }{32}}   (8+4g+9g^2 +\mathcal{O}(g^3 ) )  & {\rm for}\ {\rm arg}(g) =0 \cr
  \sqrt{\frac{\pi }{32}}   (8+4g+9g^2 +\mathcal{O}(g^3 ) ) 
-i \sqrt{\frac{\pi }{32}}    e^{-\frac{1}{2 g}} (8-4g+9g^2  +\mathcal{O}(g^3 )) 
& {\rm for}\ {\rm arg}(g) \in (0,\pi ) \cr
 \sqrt{\frac{\pi }{32}}   (8+4g+9g^2 +\mathcal{O}(g^3 ) ) 
+i \sqrt{\frac{\pi }{32}}    e^{-\frac{1}{2 g}} (8-4g+9g^2  +\mathcal{O}(g^3 )) & {\rm for}\ {\rm arg}(g) \in (-\pi ,0 ) \cr
\end{matrix}\right. ,
\end{\eq}
while large-$g$ expansion is given by
\begin{\eq}
F(g)
=\pi g^{-1/2}\left( 1 -\frac{1}{4}  g^{-1} +\frac{3}{64}  g^{-2}   -\frac{5}{768} g^{-3}
  +\mathcal{O}(g^{-4})  \right) .
\end{\eq}
Notice that the small-$g$ expansion has a Stokes line at
arg$(g)=0,\ \pi$ and anti-Stokes lines at arg$(g) = \pm\pi/2$ as
summarized in fig.~\ref{fig:toymodel_0d_stokes}.  We can again
understand this from the viewpoint of saddle points analysis.  Saddle
points of the integration are given by $x_\ast = 0, \pm \pi/2$.  At
the saddle points, the action $S(x)=\frac{1}{2g}\sin^2{x}$ takes the
values
\begin{\eq}
S(x_\ast =0 ) =0,\quad S(x_\ast =\pm\pi/2 ) =\frac{1}{2g} .
\end{\eq}
We can pick up all the saddle points through steepest descent
except\footnote{ Note that for ${\rm arg}(g)=0_\pm$, direction of the
  steepest descent around $x_\ast =\pm \pi/2$ is
  ${\rm arg}(x)=\mp (\pi /2 -0_\pm )$ while the one for $x_\ast =0$ is
  ${\rm arg}(g)=0_\pm$.  } for ${\rm arg}(g)=0$.  We again have
relative minus sign in the contributions from the non-trivial saddle
points $x_\ast =\pm\pi/2$ because directions of the steepest descent
through $x_\ast =\pm\pi /2$ are opposite between the cases for
$0 < {\rm arg}(g)<\pi$ and $-\pi < {\rm arg}(g)< 0$.
As in the 0d $\varphi^4$ theory, below we consider interpolating
functions along the positive and negative real axis, and study their
analytic properties as complex functions.

\begin{figure}[t]
\begin{center}
\includegraphics[width=7.4cm]{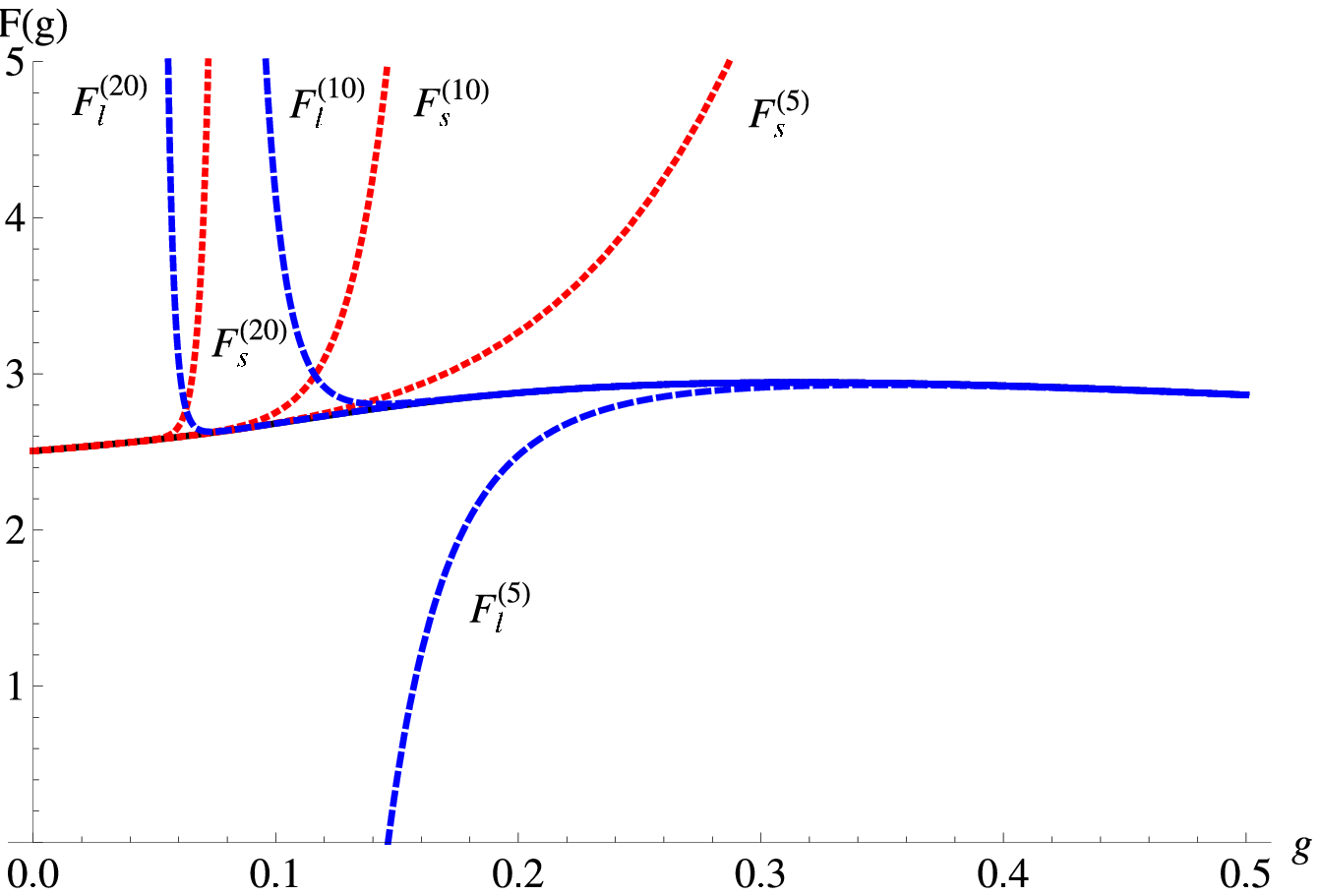}
\includegraphics[width=7.4cm]{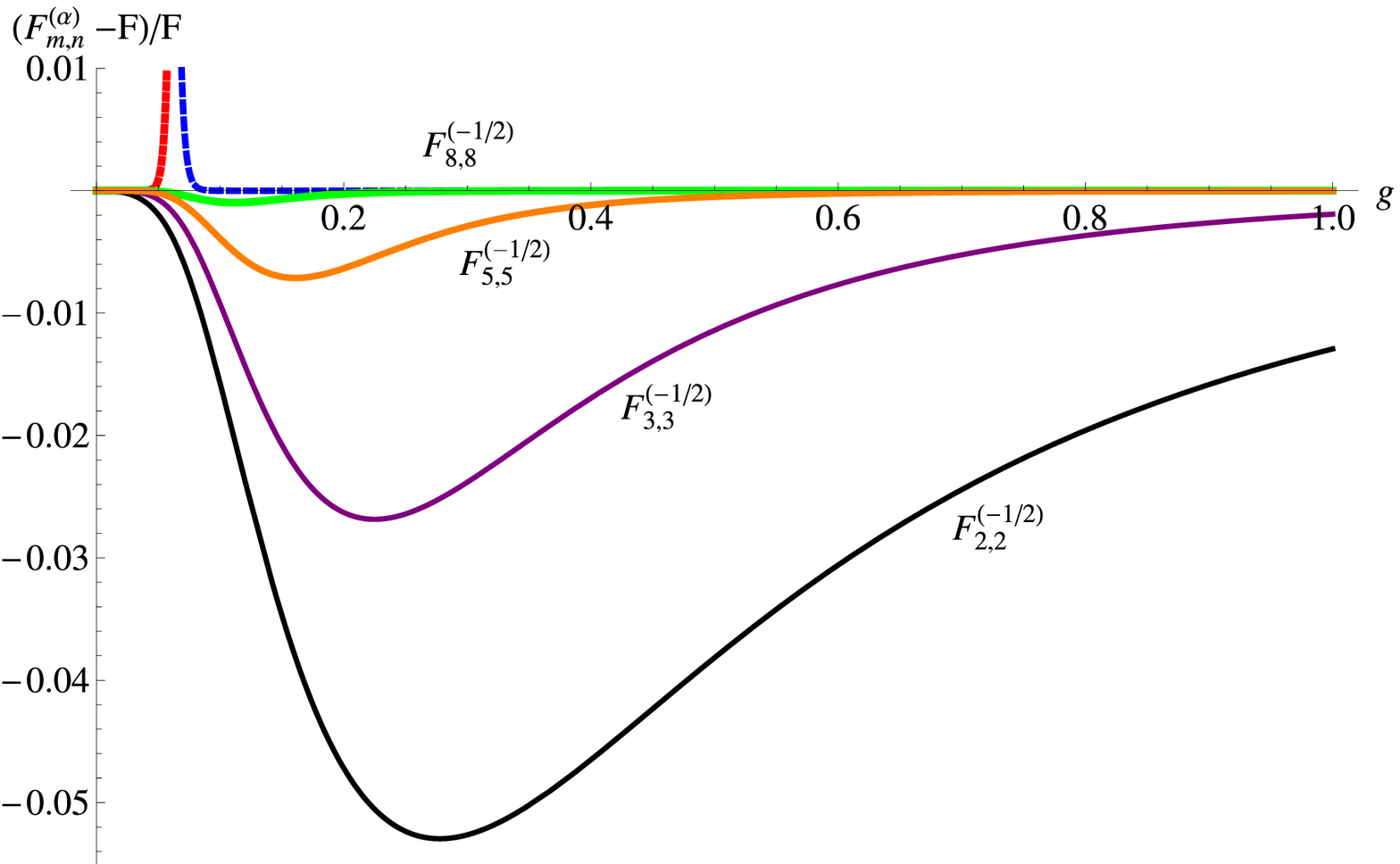}
\end{center}
\caption{ [Left] The partition function of the 0d Sine-Gordon model
  (black solid), its small-$g$ expansions $F_s^{(N_s)}(g)$ (red
  dotted) and large-$g$ expansions $F_l^{(N_l)}(g)$ (blue dashed) for
  ${\rm arg}(g)=0$.  [Right] The function $(F_{m,n}^{(\alpha )}/F -1)$
  is plotted to $g$ for some $(m,n,\alpha )$.}
\label{fig:toymodel_realp}
\end{figure}
\subsection{Interpolation along positive real axis}
We start with interpolating functions along the non-negative real axis
of $g$ and then analytically continue to complex coupling.  For this
case, we have the following small-$g$ and large-$g$ expansions
\begin{\eq}
F_s^{(N_s )}(g) = \sum_{k=0}^{N_s} s_k g^k ,\quad 
s_k=\sqrt{2\pi} \frac{2^k \Gamma^2 (k+1/2) }{\Gamma (k+1) 
\Gamma^2 (1/2)}   ,\NN
\end{\eq}
\begin{\eq}
F_{l}^{(N_l )}(g) = g^{-1/2}\sum_{l=0}^{N_l} l_k g^{-k} ,\quad 
l_k = \sqrt{\pi}\frac{\Gamma (k+1/2)}{\Gamma (k+1)} ,
\end{\eq}
which are compared with the exact result in
fig.~\ref{fig:toymodel_realp} [Left].  In terms of these expansions,
we can construct interpolating function $F_{m,n}^{(\alpha )}(g)$ (see
app.~\ref{app:toymodel} for explicit forms).

\begin{figure}[t]
\begin{center}
\includegraphics[width=7.3cm]{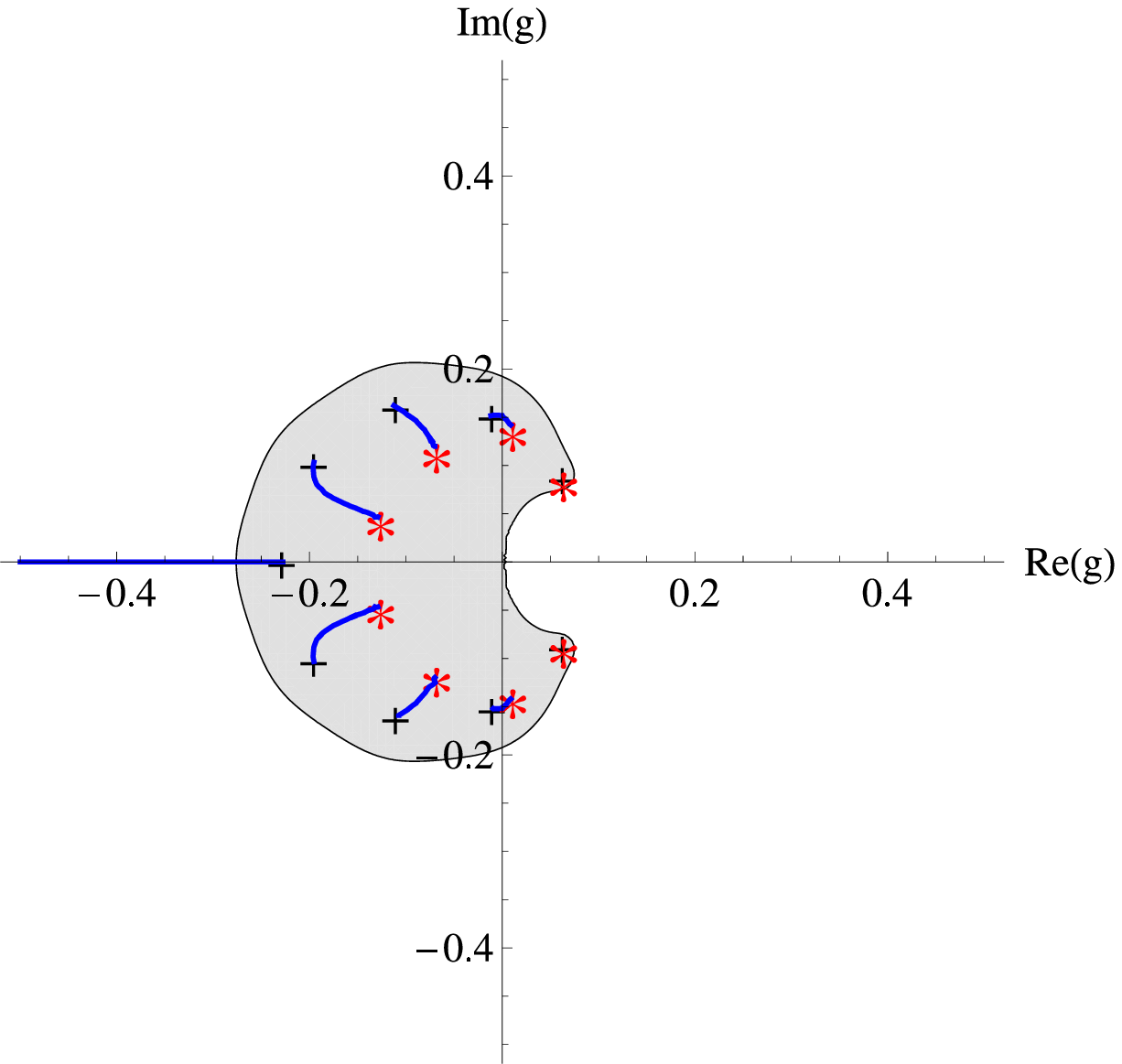}
\includegraphics[width=8.0cm]{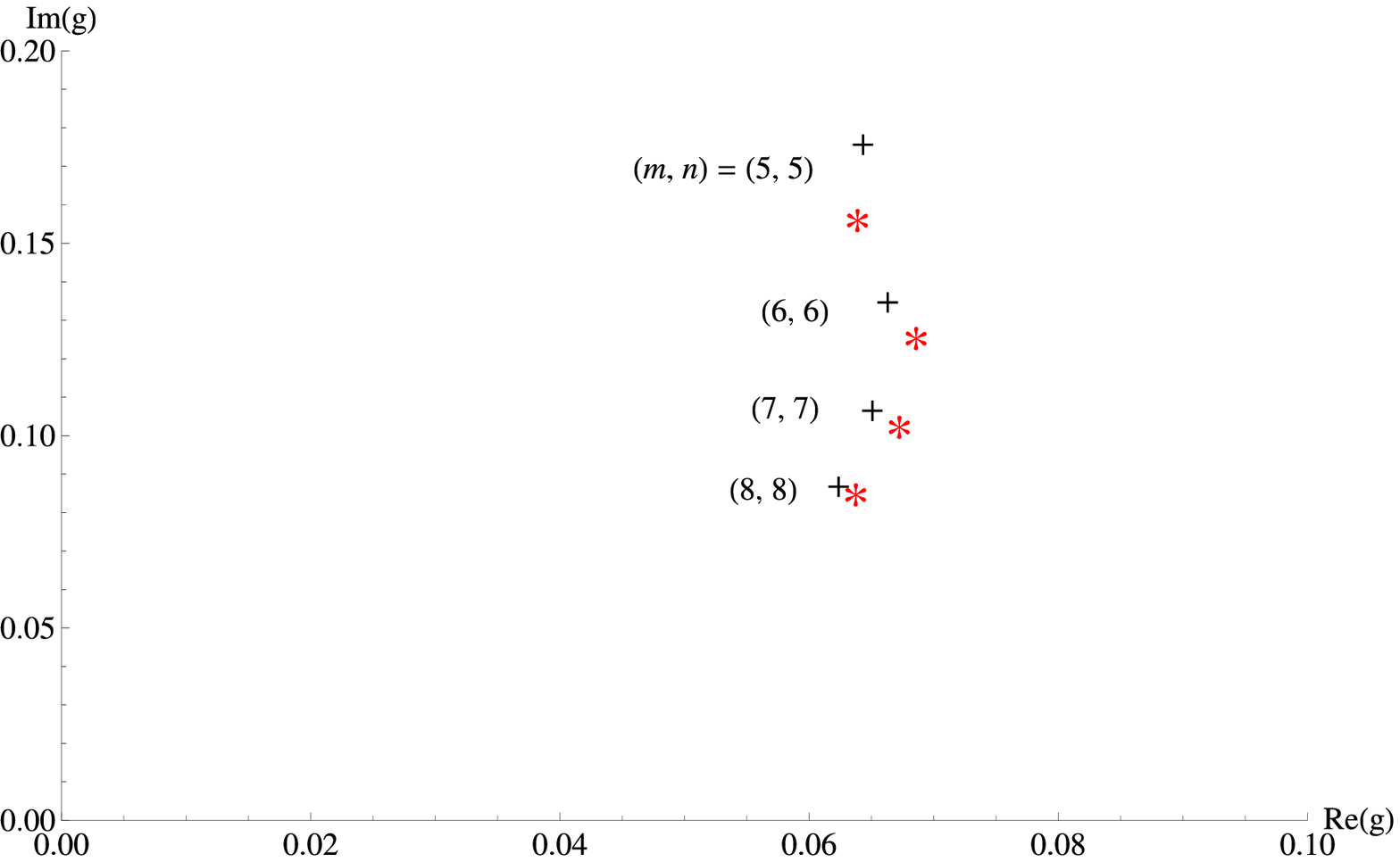}
\end{center}
\caption{ [Left] In the shaded region, the ratio $|F_{8,8}^{(-1/2)}/F-1|$ is larger
  than $0.05$.  We also plot zeros ($+$) and poles ($\ast$) of $F_{8,8}^{(-1/2)}(g)$.  The blue
  solid line denotes the branch cuts.  [Right] First poles and zeros
  of $(F_{m,n}^{(-1/2 )})^2$ from the positive real axis for various $(m,n)$.  }
\label{fig:toymodel_rp_complex}
\end{figure}

In fig.~\ref{fig:toymodel_realp} [Right], we compare some FPRs with the exact result for ${\rm arg}(g)=0$.  
As in the 0d $\varphi^4$ theory, we find that these interpolating functions well
approximate the exact result for positive real $g$.  For example, the
best approximation $F_{8,8}^{(-1/2)}(g)$ among the FPRs has
$\mathcal{O}(0.1\% )$ error at worst.  Next let us consider complex
$g$ in the best FPR $F_{8,8}^{(-1/2)}(g)$.  We summarize the validity
of the approximation with analytic structures of $F_{8,8}^{(-1/2)}(g)$
in fig.~\ref{fig:toymodel_rp_complex} [Left].  
The shaded region again starts with the first branch cuts seen from the positive real axis, 
which are given by the first poles and zeros of the rational function $(F_{8,8}^{(-1/2)}(g))^2$.  
The first poles are located at
${\rm arg}(g)\simeq \pm 0.9348$, which is not near from the
anti-Stokes lines at ${\rm arg}(g)=\pi /2 \simeq \pm 1.571$.  We again
expect that the first branch cuts will vanish as increasing $(m,n)$ in $F_{m,n}^{(-1/2)}(g)$.  
Indeed fig.~\ref{fig:toymodel_rp_complex}
[Right] implies that the first poles will collapse to the first zeros
for large $(m,n)$ as in sec.~\ref{sec:phi4_negative}.
Hence we conclude that the
branch cuts are the fake singularities and the FPR
$F_{m,n}^{(-1/2)}(g)$ with sufficiently large $(m,n)$ would give good
approximation in the right-half plane of $g$.  In next subsection, we
aim to construct interpolating functions approximating the exact
result on the left-half plane by considering interpolating problem
along the negative real axis.

\subsection{Interpolation along negative real axis}
\begin{figure}[t]
\begin{center}
\includegraphics[width=7.4cm]{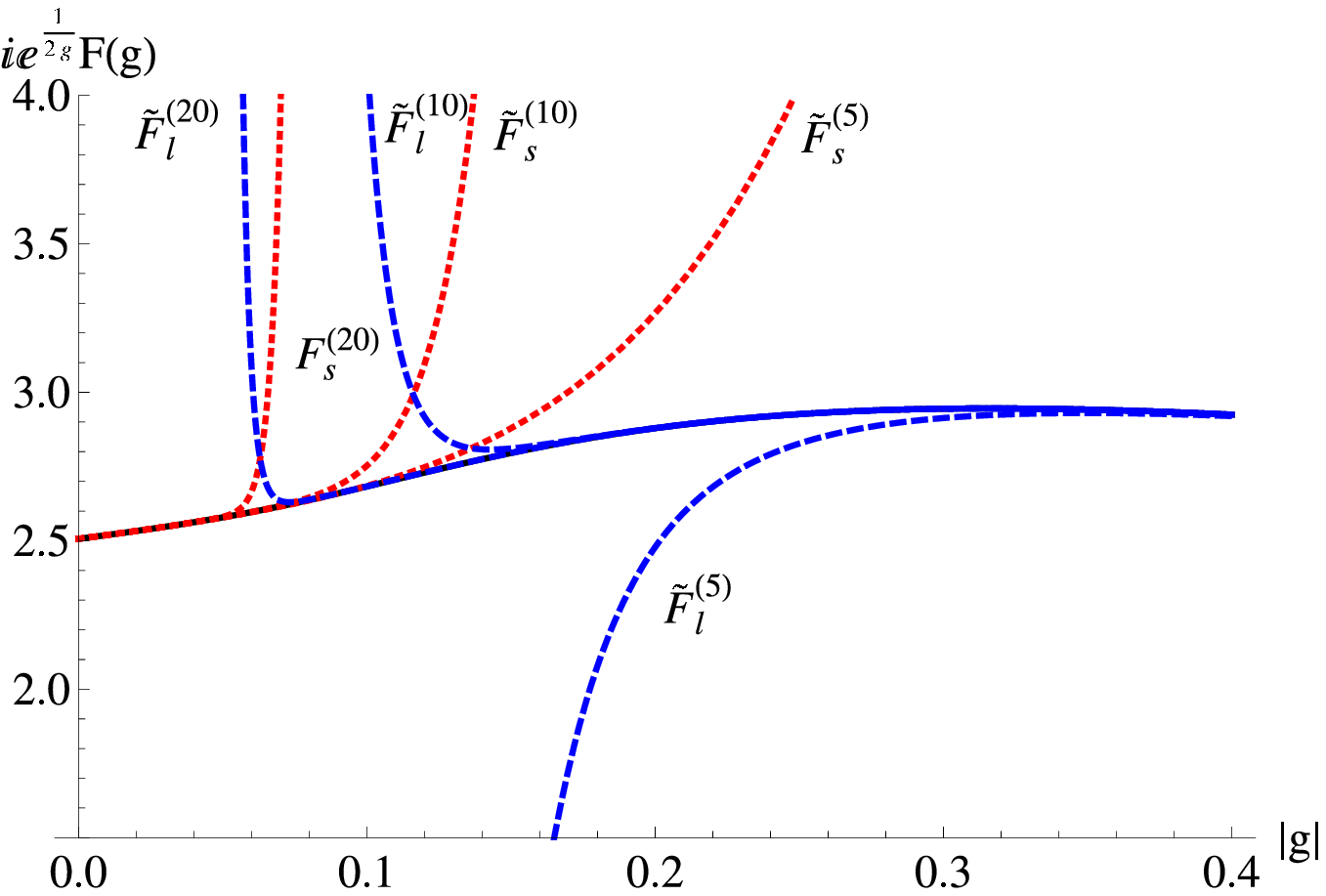}
\includegraphics[width=7.4cm]{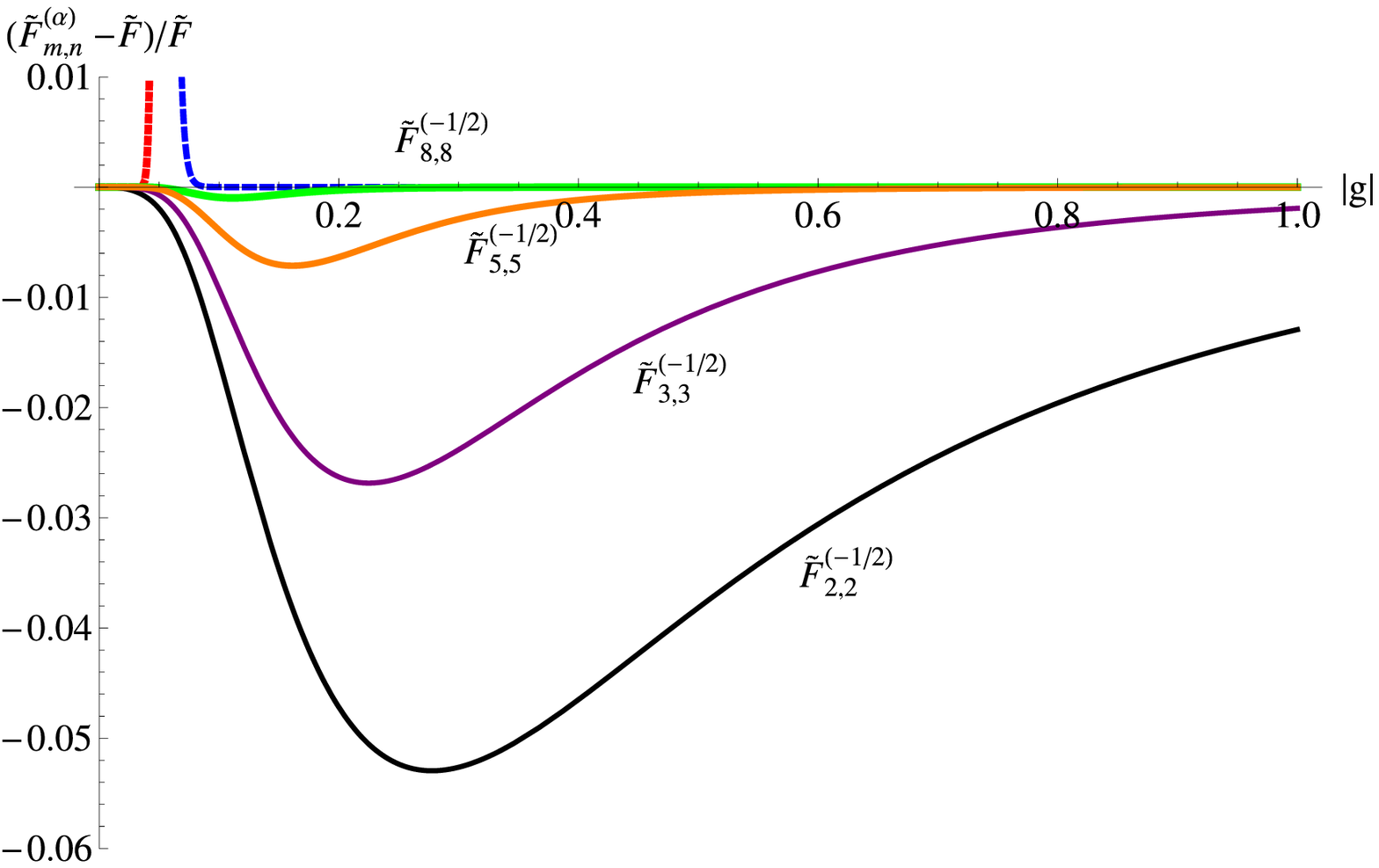}
\end{center}
\caption{ [Left] The function $\tilde{F}(g)=ie^{\frac{1}{2g}}F(g)$
  (black solid), its small-$g$ expansions $\tilde{F}_s^{(N_s)}(g)$
  (red dotted) and large-$g$ expansions $\tilde{F}_l^{(N_l)}(g)$ (blue
  dashed) for ${\rm arg}(g)=\pi -\epsilon$. 
[Right] The relative error $(\tilde{F}_{m,n}^{(\alpha )}/\tilde{F}-1)$ is plotted to $|g|$ for real negative $g$.}
\label{fig:toymodel_realn}
\end{figure}
Let us consider interpolating problem along the negative real axis.
The function $F(g)$ has the following expansions
\begin{\eqa}
F(-t+i\epsilon )
&=& \sqrt{\frac{\pi }{32}}   (8-4t+9t^2 +\mathcal{O}(t^3 ) ) 
 -i \sqrt{\frac{\pi }{32}}  e^{\frac{1}{2t}} (8+4t-9t^2  +\mathcal{O}(t^3 )) +\mathcal{O}(\epsilon ) \NN\\
&=&-i t^{-1/2} \left(  \pi +\frac{1}{4}\pi t^{-1}  +\frac{3}{64}\pi t^{-2}  +\frac{5}{768}  \pi t^{-3} +\mathcal{O}(t^{-4}) \right)
 +\mathcal{O}(\epsilon ) , \NN\\
F(-t-i\epsilon )
&=&\sqrt{\frac{\pi }{32}}   (8-4t+9t^2 +\mathcal{O}(t^3 ) ) 
 +i \sqrt{\frac{\pi }{32}}  e^{\frac{1}{2t}} (8+4t-9t^2  +\mathcal{O}(t^3 )) 
 +\mathcal{O}(\epsilon ) ,\NN\\
&=&+i t^{-1/2} \left(  \pi +\frac{1}{4}\pi t^{-1}  +\frac{3}{64}\pi t^{-2}  +\frac{5}{768}  \pi t^{-3} +\mathcal{O}(t^{-4}) \right)
 +\mathcal{O}(\epsilon ) ,
\end{\eqa}
with $t\in\mathbb{R}_+$.  The dominant parts of the expansions change
the signs across the negative real axis as in the $\varphi^4$ theory
since $F(g)$ has the branch cut on the real negative axis.  Hence it
is more appropriate to consider the function
\begin{\eq}
\tilde{F}(t) =\lim_{\epsilon\rightarrow +0}
\left.  i e^{\frac{1}{2g}} F(g+i\epsilon ) \right|_{g\rightarrow -t}, 
\quad t\in\mathbb{R}_+ .
\end{\eq}
The function $\tilde{F}(t)$ has the expansions,
\begin{\eqa}
\tilde{F}(t)
&=&\sqrt{2 \pi } +\sqrt{\frac{\pi }{2}} t +\frac{9}{4} \sqrt{\frac{\pi }{2}} t^2 +\frac{75}{8} \sqrt{\frac{\pi }{2}} t^3 +\mathcal{O}(t^4) \NN\\
&=& t^{-1/2}\left( \pi  -\frac{1}{4} \pi t^{-1} +\frac{3}{64} \pi t^{-2} -\frac{5}{768} \pi t^{-3} +\mathcal{O}(t^{-4}) \right) ,
\end{\eqa}
where we have dropped the exponentially suppressed correction
$\mathcal{O}(e^{-\frac{1}{2t}})$ in the small-$g$ expansion (see
fig.~\ref{fig:toymodel_realn} [Left] for comparison of these
expansions with the exact result of $\tilde{F}(t)$).  Then we can
construct the FPR $\tilde{F}_{m,n}^{(\alpha )}(t)$ to interpolate
these expansions and approximate the original function $F(g)$ by
\begin{\eq}
F_{L\pm ,m,n}^{(\alpha )} (g) =  \mp i e^{-\frac{1}{2g}} \tilde{F}_{m,n}^{(\alpha )} (-g) .
\end{\eq}
The function $F_{L\pm ,m,n}^{(\alpha )} (g) $ reproduces the small-$g$
and large-$g$ expansions of $F(g)$ for
$g\in \mathbb{R}_- \pm i\epsilon$ up to certain orders.

\begin{figure}[t]
\begin{center}
\includegraphics[width=7.4cm]{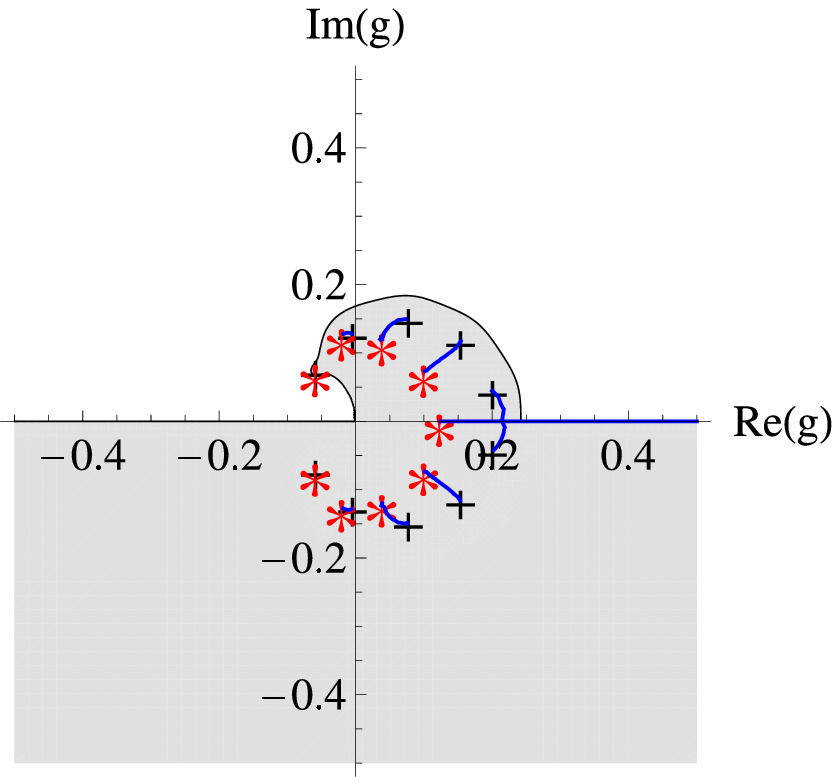}
\includegraphics[width=7.4cm]{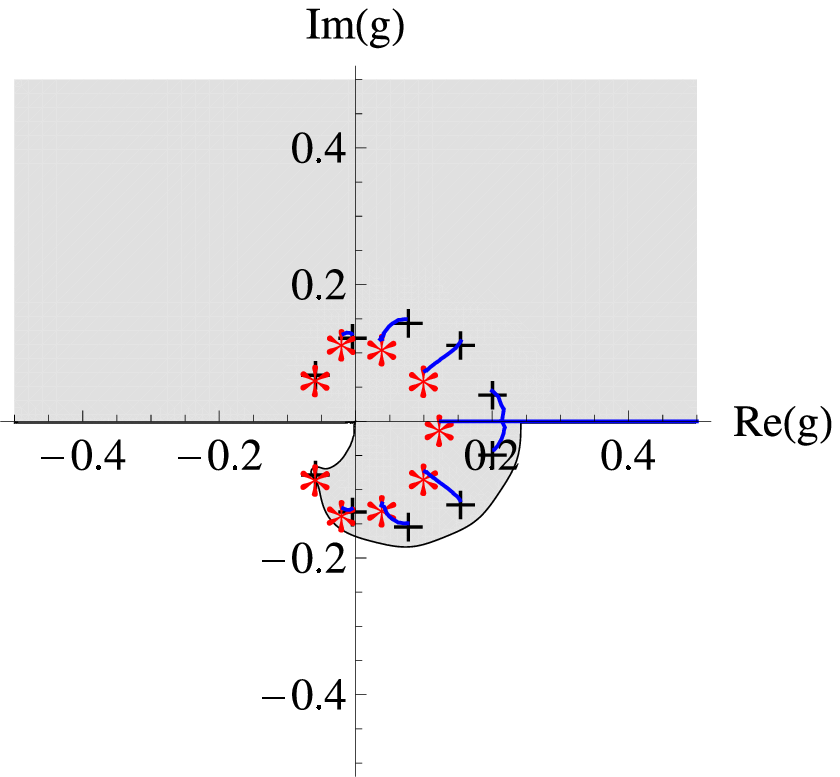}
\end{center}
\caption{[Left] The region where the ratio $|F_{L+,9,9}^{(-1/2)}/F-1|$
  is larger than $0.05$.  We also plot zeros ($+$)
  and poles ($\ast$) of the rational function $(\tilde{F}_{10,10}^{(1/2)})^2$.  
 The blue solid lines denote the
  branch cuts of $F_{L+,9,9}^{(-1/2)}$.  
  [Right] Similar plot as the left for $F_{L-,9,9}^{(-1/2 )}$.  }
\label{fig:toymodel_rn_complex}
\end{figure}

\begin{figure}[tbp]
\begin{center}
\includegraphics[width=8.0cm]{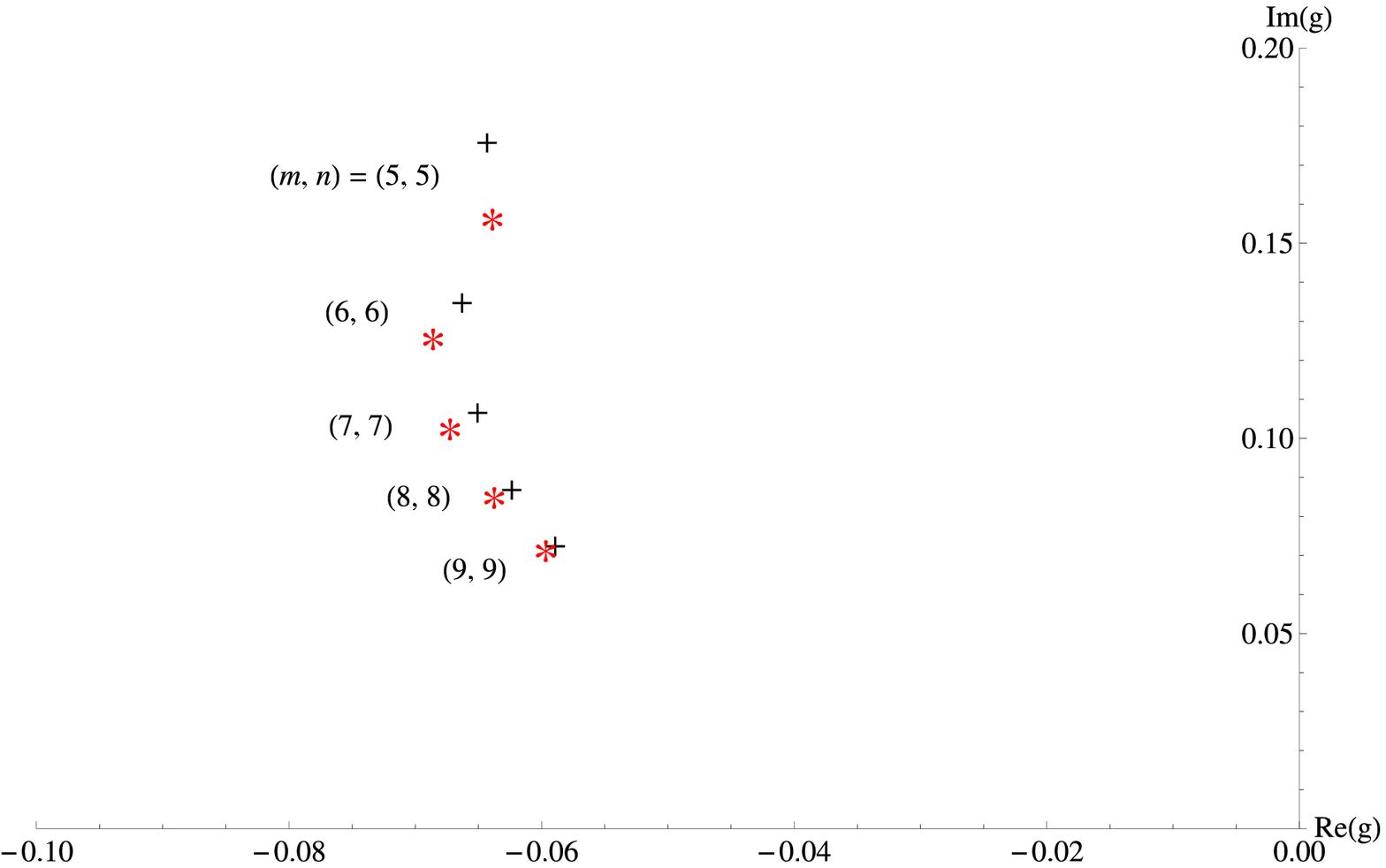}
\includegraphics[width=7.4cm]{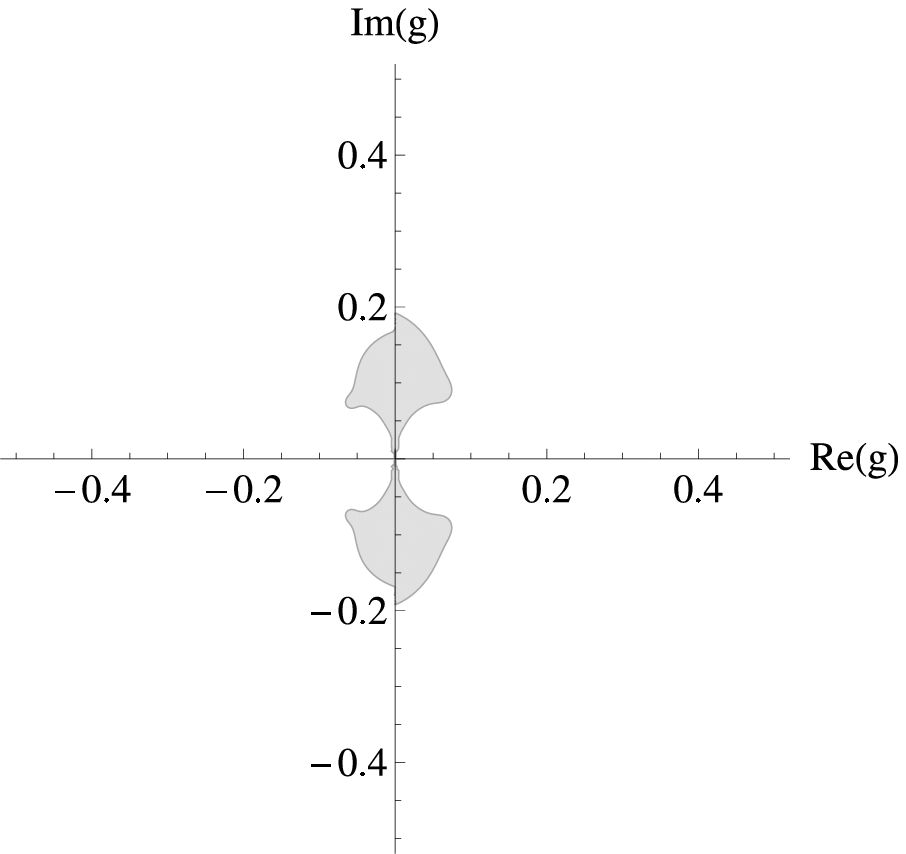}
\end{center}
\caption{ [Left] First poles and zeros of $(F_{L+,m,n}^{(-1/2 )})^2$
  from the positive real axis for various $(m,n)$. 
  [Right] The region where the patch $F'(g)$ of the best FPRs gives more than $5\%$ error.  }
\label{fig:toymodel_1st_poles}
\end{figure}

In fig.~\ref{fig:toymodel_realn} [Right] we check that
the interpolating functions $\tilde{F}_{m,n}^{(\alpha )} (g) $ well approximate $\tilde{F}(g)$ on the negative real axis.  
We again summarize the
validity of approximation by $F_{L\pm ,9,9}^{(-1/2 )} (g) $ and its
analytic property in fig.~\ref{fig:toymodel_rn_complex}.  Starting
with the negative real axis counterclockwise (clockwise), the
approximation $F_{L+ ,9,9}^{(-1/2 )} (g) (F_{L- ,9,9}^{(-1/2 )} (g)) $
gets worse across the first branch cut as in last subsection, which is
located around the line ${\rm arg}(g)\simeq (-)2.2578$ and deviates from the anti-Stokes lines.
However, according to fig.~\ref{fig:toymodel_1st_poles} [Left], the
first branch cuts seems to shrink as increasing $(m,n)$ in
$(F_{L\pm m,n}^{(-1/2)}(g))^2$.  Thus we expect that the branch cut is
the fake singularity and the FPR $F_{L+, m,n}^{(-1/2)}(g)$
($F_{L-, m,n}^{(-1/2)}(g)$) with sufficiently large $(m,n)$ well
approximates $F(g)$ in the left-top-quarter (left-bottom-quarter)
plane.

Let us patch the best interpolating functions along the positive and
negative real axis as in the $\varphi^4$ theory.  In
fig,~\ref{fig:toymodel_1st_poles} [Right], we draw range of validity
of approximation by
\begin{\eq}
F' (g) = \left\{ \begin{matrix}
F_{8,8}^{(-1/2)} (g)        & {\rm for}& |{\rm arg}(g)| <\pi /2 \cr
F_{L+,9,9}^{(-1/2)} (g) & {\rm for}& \pi /2 < {\rm arg}(g) <\pi \cr
F_{L-,9,9}^{(-1/2)} (g) & {\rm for}& -\pi  < {\rm arg}(g) < -\pi /2 \cr
\end{matrix}\right. .
\end{\eq}
This indicates that the patching $F' (g)$ has $5\%$ or better accuracy
in the fairly wide region.

\section{Conclusion and Discussions}
\label{sec:conclusion}
We have studied analytic structures of some interpolating functions
and discussed their physical implications.  We have proposed that the
analytic structures of the interpolating functions provide information
on analytic property and Stokes phenomena of the physical quantity,
which we approximate by the interpolating functions.  More concretely,
we have mainly considered the roles of the first branch cuts measured
from a specific axis where we construct the interpolating
functions.  If the first branch cuts are not ``fake singularities",
then we expect that the cuts approximate those of the exact result or
indicate locations of Stokes or anti-Stokes lines.  When the cuts are
fake singularities, we should consider next cuts in order to get
physical implications.  We have explicitly checked our proposal in the
partition functions of the 0d $\varphi^4$ theory and the Sine-Gordon
model.  This seems to hold also in other examples such as BPS Wilson
loop in 4d $\mathcal{N}=4$ Super Yang-Mills theory, energy spectrum in
1d anharmonic oscillator etc \cite{progress}.

We expect that our result is applicable in more practical problems,
where we do not know exact solutions.  One possible application of
our results is to use them to find Stokes behavior
of the physical quantity by studying analytic structures of 
interpolating functions.

We have compared the result of the FPR with the recent result \cite{Cherman:2014xia} of
the resurgence in the 0d Sine-Gordon model.  We have seen that the finite
order approximation of the resurgence give better approximations than
the FPR in the region where the FPR gives relatively imprecise
approximation while the FPR is more precise in sufficiently strong
coupling region.  This implies that the FPR and resurgence play
complementary roles.  It will be interesting to compare them in more
detail.

One of subtle points in our approach is concerned with the fake
singularities.  It is unclear how we should extend the definition of
the fake singularities to more general problem.  We have defined the
fake singularity in our two examples as the branch cut shrinking for
large $(m,n)$ with fixed $\alpha$.  Since the relative errors seem to
monotonically decrease as increasing $(m,n)$ in our examples, it
would be reasonable to expect that the FPRs with sufficiently large
$(m,n)$ have larger range of validity in the complex $g$-plane.
However, in general problem, FPRs with larger $(m,n)$ do not
necessarily give better approximations as seen in
\cite{Honda:2014bza}.  Natural extension of the definition would be to
consider a family of FPRs $\{ F_{m_s ,n_s}^{(\alpha_s )} \}$
satisfying
\begin{\eq}
\left| \frac{F_{m_s ,n_s}^{(\alpha_s )} }{F(g)} -1 \right|  <\left| \frac{F_{m_{s+1} ,n_{s+1}}^{(\alpha_{s+1})} }{F(g)} -1 \right| ,
\end{\eq}
and define fake singularity as a cut, which shrinks as we increase
$s$.  It would be interesting to pursue this direction further.

In this paper we have focused on analytic properties of the best
interpolating functions, which provide the best approximation among
the given interpolating functions along a specific axis.  However,
when we do not know the exact results, it is nontrivial which
interpolating functions give relatively better approximation as
discussed in \cite{Honda:2014bza}.  Although the work
\cite{Honda:2014bza} has proposed a criterion to choose the best
interpolating function in terms of two perturbative expansions, we
need information on large order behavior of the expansions to use the
criterion.  It is nice if we use analytic properties of the
interpolating functions to determine the relatively better interpolating
functions without knowing exact results.  In our examples, the exact
results have the square-root type branch cuts.  Interestingly the best
interpolating functions are at $\alpha =\pm 1/2$ and hence also have the
square-root type branch cuts.  Thus the analytic properties of the
interpolating functions would be helpful to improve such criterion.

We have seen that each FPR considered here has its own angular wedge
of validity.  By patching the best FPRs along the positive and
negative real axis, we have obtained the approximation with larger
range of validity than each FPR.  It would be interesting if one can
construct single interpolating function, which gives small-$g$ and
large-$g$ expansions for all ${\rm arg}(g)$.  One might think that
this was conceptually similar to finding connection formula between
different Stokes domains in exact JWKB method.  However there are some
important differences.  One of such differences is that FPR often
still gives good approximation even across Stokes line while
approximation by FPR necessarily breaks down across the anti-Stokes line.  
For example, in the 0d $\varphi^4$ theory, the best FPR along
the positive real axis gives precise approximation even across the Stokes
line at ${\rm arg}(g)=\pi /2$ unless we approach to the anti-Stokes
line at ${\rm arg}(g)=3\pi /4$.  Thus the connection formula in JWKB
does not seem to give useful hints.  Let us see one of main
difficulties to construct the single interpolating function valid for
all ${\rm arg}(g)$ in the toy example:
\[
F(g) =F_0 (g) +e^{-\frac{1}{g}} F_1 (g) ,\quad {\rm with}\ 
F_0 (g) = \sum_{k=0}^\infty a_{0,k} g^k  ,\ F_1 (g) = \sum_{k=0}^\infty a_{1,k} g^k ,
\]
where the summations $F_0 (g)$ and $F_1 (g)$ are convergent, and the
exact result of $F(g)$ is given by their analytic continuations.  Let
us consider FPR to interpolate small-$g$ and large-$g$ expansions of
$F(g)$.  If we construct FPR along the positive real axis, then the
FPR interpolates $F_0 (g)$ and the large-$g$ expansion of $F(g)$.
Some part of the information about the large-$g$ expansion is encoded
in $F_1 (g)$, but the FPR does not have this information in small-$g$ regime.
This missing information gives exponentially suppressed error on the
right plane of $g$ and exponentially growing error on the left plane
for small $|g|$.  Similarly if we construct FPR along the negative
real axis, then that FPR does not have the information that a part of
the large-$g$ expansion comes from $F_0 (g)$.  
Of course we can find much better approximation in this example by separately constructing FPRs to
interpolate $F_{0,1} (g)$ and large-$g$ expansion of analytic
continuation of $F_{0,1} (g)$.  However this information is almost
equivalent to have the exact result and there is no
motivation to perform this procedure.

\subsection*{Acknowledgments} We would like to thank Ashoke Sen for
discussions and useful suggestions.  DPJ would like to thank IMSc,
Chennai and CHEP, Bengaluru for hospitality during the course of this
work.  This work was supported in parts by the DAE project 12-R\&
D-HRI-5.02.-0303.

\appendix
\section{Comparison with resurgence result}
\label{app:resurgence}
As mentioned in the main text, the authors in \cite{Cherman:2014xia} have
performed resurgence analysis in the 0d Sine-Gordon model.  It is interesting to
compare their result with our interpolating functions.  Their result
is
\begin{\eq}
F(g)
=\left\{ \begin{matrix}
\mathcal{S}_{{\rm arg}(g)} \Phi_0 (g) -ie^{-\frac{1}{2g}} \mathcal{S}_{{\rm arg}(g)}\Phi_1 (g)
& {\rm for}\ {\rm arg}(g)\in (0,\pi ) \cr
\mathcal{S}_{{\rm arg}(g)} \Phi_0 (g) +ie^{-\frac{1}{2g}} \mathcal{S}_{{\rm arg}(g)}\Phi_1 (g)
& {\rm for}\ {\rm arg}(g)\in (-\pi ,0 ) \cr
\end{matrix}\right. ,
\label{eq:resurgence}
\end{\eq}
where
\begin{\eq}
\mathcal{S}_\theta \Phi_{0,1}(g)
=\frac{1}{g}\int_0^{+\infty e^{i\theta}} dt\ 
e^{-\frac{t}{g}} \tilde{\mathcal{B}}\Phi_{0,1}(t) .
\label{eq:comeback}
\end{\eq}
The function $\tilde{\mathcal{B}}\Phi_{0}(t)$
($\tilde{\mathcal{B}}\Phi_{1}(t)$) denotes analytic continuation of
Borel transformation of the small-$g$ expansion coming from the saddle
point $x_\ast =0$ ($x_\ast =\pm\pi/2$), which is explicitly given by
\begin{\eq}
\tilde{\mathcal{B}}\Phi_{0}(t) 
=\sqrt{2\pi}\ _2 F_1 \left( \frac{1}{2},\frac{1}{2};1;2t \right) ,\quad
\tilde{\mathcal{B}}\Phi_{1}(t)
=\sqrt{2\pi}\ _2 F_1 \left( \frac{1}{2},\frac{1}{2};1;-2t \right) .
\end{\eq}
Although we know all the coefficients of the small-$t$ expansions, let
us consider finite order approximation of the resurgence result as in
\cite{Cherman:2014xia} and compare this with the FPRs.  Namely,
instead of using $\tilde{\mathcal{B}}\Phi_{0,1}(t)$, we terminate its
small-$t$ expansion up to $\mathcal{O}(t^{2N+1})$ and use its
one-point Pad\'e approximant\footnote{ Strictly speaking, this is
  so-called diagonal Pad\'e approximant.  }:
\begin{\eq}
P^{(2N+1)} (t) = \frac{\sum_{k=0}^N c_k t^k}{1+\sum_{k=1}^N d_k t^k} ,
\end{\eq}
in the integration \eqref{eq:comeback}, 
which reproduces the small-$t$ expansion up to $\mathcal{O}(t^{2N+1})$.
This procedure is often called (one-point) Borel-Pad\'e approximation.

\begin{figure}[tbp]
\begin{center}
\includegraphics[width=7.4cm]{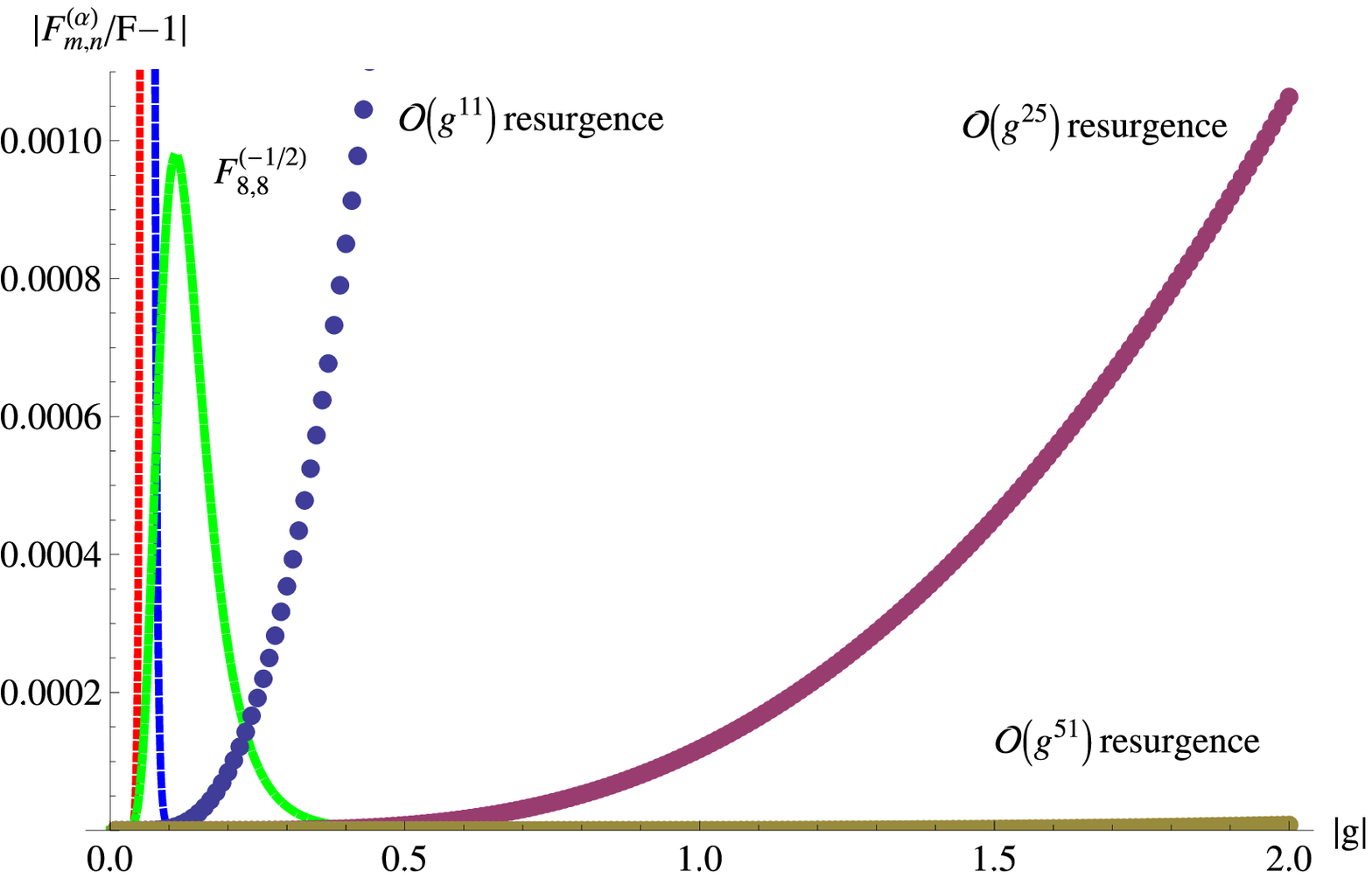}
\includegraphics[width=7.4cm]{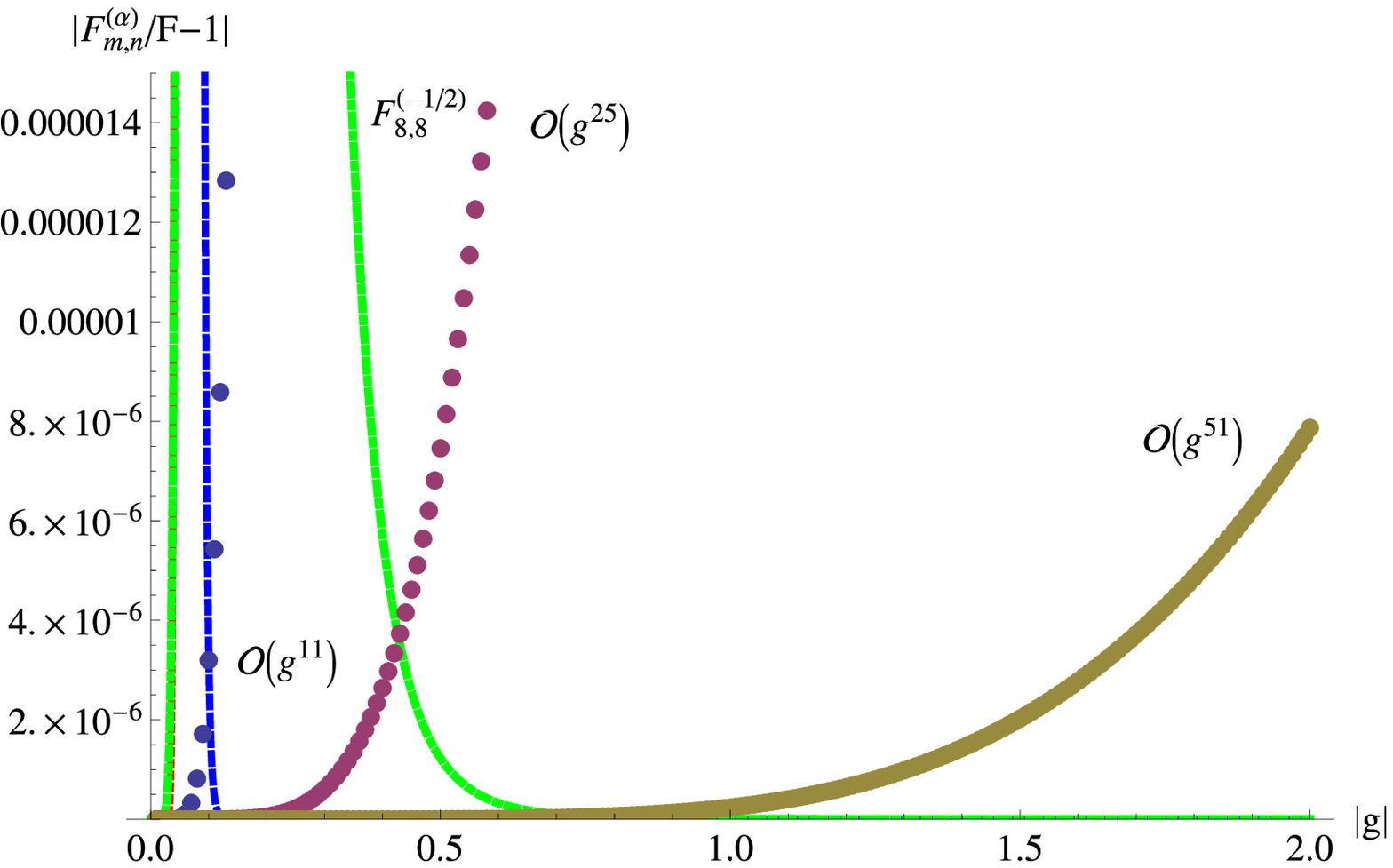}
\end{center}
\caption{[Left] $|F_{8,8}^{(-1/2 )} /F-1 |$ for ${\rm arg}(g)=\frac{\pi}{100} $
[Right] The same plot as the left in a different scale.  }
\label{fig:resurgence}
\end{figure}
In fig.~\ref{fig:resurgence}, we compare the result of the resurgence
with the FPR $F_{8,8}^{(-1/2)}$ for\footnote{ This choice of
  ${\rm arg}(g)$ is due to a technical problem in the integration
  \eqref{eq:comeback}.  As decreasing ${\rm arg}(g)$, it becomes
  harder to obtain precise values of the integration.  Here we expect
  that this result for ${\rm arg}(g)=\frac{\pi}{100}$ is almost the
  same as for ${\rm arg}(g)\rightarrow0$.  }
${\rm arg}(g)=\frac{\pi}{100}$.  By ``$\mathcal{O}(g^M )$ resurgence",
we mean the result obtained by the replacement
$\tilde{\mathcal{B}}\Phi_{0,1}(t) \rightarrow P^{(M)} (t)$ in
\eqref{eq:comeback} and \eqref{eq:resurgence}.  We first find that the
resurgence result gives very precise approximation in weak coupling regime.  
Furthermore 
all the results of the resurgence give better approximations 
in the region where the FPR gives relatively imprecise approximation. 
As we go to the large-$|g|$ regime, the approximation
by the resurgence becomes monotonically worse but the
$\mathcal{O}(g^{51})$ resurgence still gives reasonable approximation at $|g|=2$.  
However, if we further go to very large-$|g|$
regime, the approximation will breakdown\footnote{ Of course, if we
  include arbitrarily higher order terms, then the resurgence can provide
  arbitrarily precise approximation.  This is also the main difference
  between resurgence method and our interpolating function approach.
}.  For example, the $\mathcal{O}(g^{51})$ resurgence has about $15\%$
error at $|g|=100$ and $140\%$ error at $|g|=1000$.  Hence in
sufficiently strong coupling regime, the FPR always gives the better
approximation than the finite order approximation of the resurgence.
This result implies that the FPR and resurgence play complementary
roles in describing the exact function at least in this example.  It
will be interesting to compare them in more detail and other examples.

\section{Comments on interpolation in Borel plane}
\label{sec:Borel}
When small-$g$ (large-$g$) expansion of the function $F(g)$ is
convergent, we expect that $F(g)$ is very precisely approximated by
the FPRs $F_{m,n}^{(\alpha )}$ with large $m(n)$.  How about the case
where small-$g$ expansion is asymptotic but Borel summable?  For this
case, Borel transformation of the small-$g$ expansion is convergent.
If we construct FPR-type interpolating function in the Borel plane,
then one might expect that the ``Borel-FPR" approximates the exact
result of $F(g)$ very well.  However, in this appendix, we discuss that
the Borel-FPR gives slightly worse approximation than the usual FPR at
least for the partition function of the 0d Sine-Gordon model.  We have
not understood any clear reasons behind this observation.

\subsection{Method}
Let us construct interpolating function for $F(g)$ in Borel plane.
For this purpose, it is convenient to introduce following quantities:
\begin{\eq}
\tilde{F}_s^{(N_s )} (g) 
= g^M F_s^{(N_s )} (g) = g^{\tilde{a}} \sum_{k=0}^{N_s} s_k g^k ,\quad
\tilde{F}_l^{(N_l )} (g) 
= g^M F_l^{(N_l )} (g) = g^{\tilde{b}} \sum_{k=0}^{N_l} l_k g^{-k} ,
\end{\eq}
where 
\begin{\eq}
\tilde{a}=M+a ,\quad \tilde{b} = M+b .
\end{\eq}
The parameter $M$ is a real number satisfying
\begin{\eq}
\tilde{a}\geq 1,\quad \tilde{b} \geq N_s +1 .
\label{eq:M}
\end{\eq}
The Borel transformation of the small-$g$ expansion is
\begin{\eqa}
\mathcal{B}\tilde{F}_s^{(N_s )}  (t) 
= \sum_{k=0}^{N_s} \frac{s_k}{\Gamma (\tilde{a}+k)} t^{\tilde{a}+k-1} .
\end{\eqa}
We also define action of $\mathcal{B}$ to the large-$g$ expansion as
\begin{\eqa}
\mathcal{B}\tilde{F}_l^{(N_l )}  (t) 
= \sum_{k=0}^{N_l} \frac{l_k}{\Gamma (\tilde{b}-k)} t^{\tilde{b}-k-1} .
\end{\eqa}
Then we can construct the FPR-type interpolating function
$\mathcal{B}F_{m,n}^{(\alpha )} (t)$ in Borel plane, which interpolate
$\mathcal{B}\tilde{F}_s^{(N_s )} (t) $ and
$\mathcal{B}\tilde{F}_l^{(N_l )} (t) $ up to
$\mathcal{O}(t^{\tilde{a}+N_s})$ and
$\mathcal{O}(t^{\tilde{b}-N_l -2})$, respectively:
\begin{\eq}
\mathcal{B}F_{m,n}^{(\alpha )} (t)  -\mathcal{B}\tilde{F}_s^{(N_s )}  (t) 
=\mathcal{O}(t^{\tilde{a}+N_s}) ,\quad
\mathcal{B}F_{m,n}^{(\alpha )} (t)  -\mathcal{B}\tilde{F}_l^{(N_l )}  (t)
 =\mathcal{O}(t^{\tilde{b}-N_l -2}) .
\end{\eq} 
This implies that the original function $F(g)$ is approximated by
\begin{\eq}
F_{m,n}^{B(\alpha )} = g^{-M} \int_0^\infty dt\ e^{-\frac{t}{g}} \mathcal{B}F_{m,n}^{(\alpha )} (t) ,
\end{\eq}
which we call ``Borel FPR".  Indeed one can show that the function
$F_{m,n}^{B(\alpha )}$ gives $F_s^{(N_s )} (g) $ in small-$g$ regime
and $F_l^{(N_l )} (g) $ in large-$g$ regime up to some
orders\footnote{ If $M$ did not satisfy the condition \eqref{eq:M},
  then we could not guarantee this property.  }.  One of subtle points in
this approach is that different values of $M$ give different
$F_{m,n}^{B(\alpha )}$ even if we consider the same $(m,n,\alpha )$.
In this paper, we will not study $M$-dependence and fix as $M=10$.

\subsection{Comparison with usual interpolating function in 0d Sine-Gordon model}
\begin{figure}[t]
\begin{center}
\includegraphics[width=7.4cm]{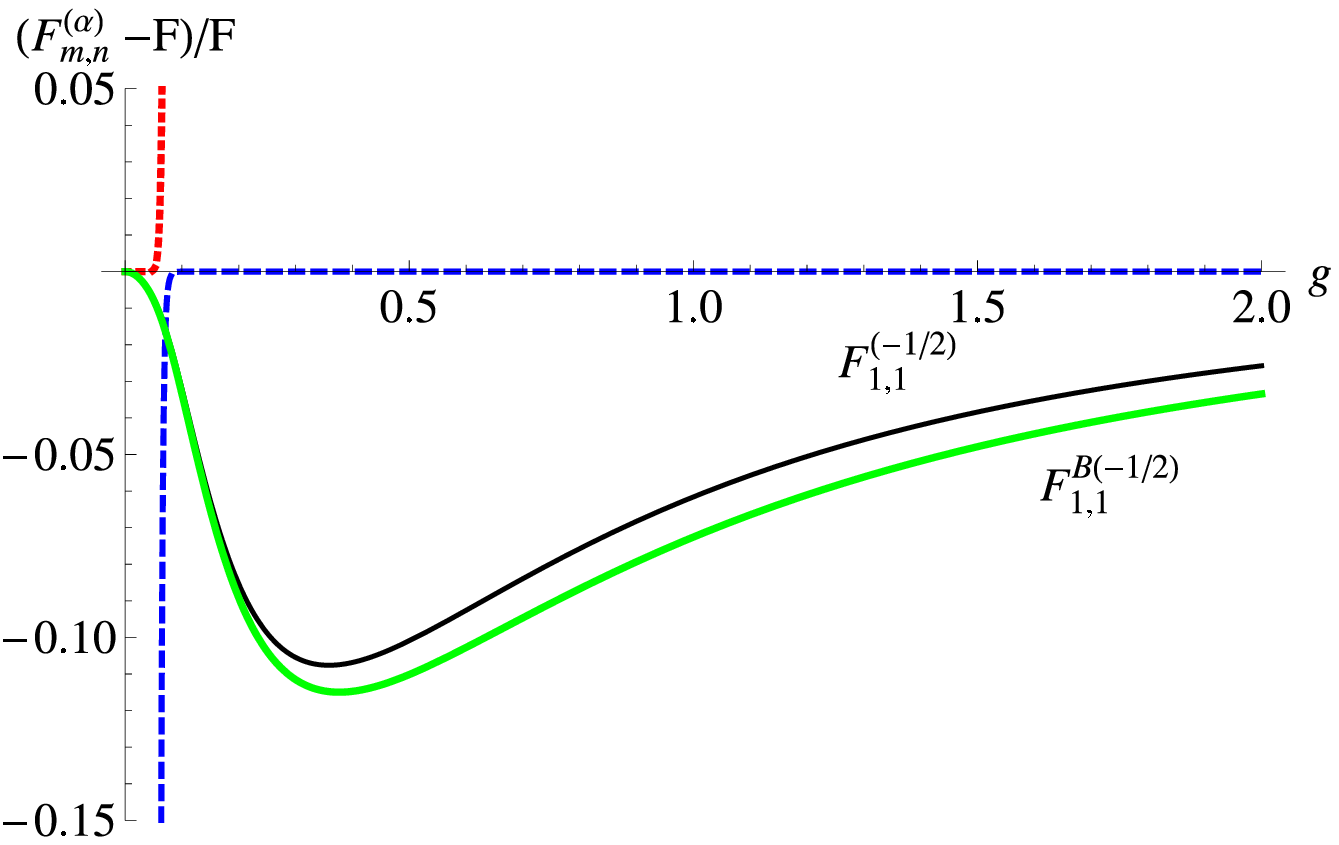}
\includegraphics[width=7.4cm]{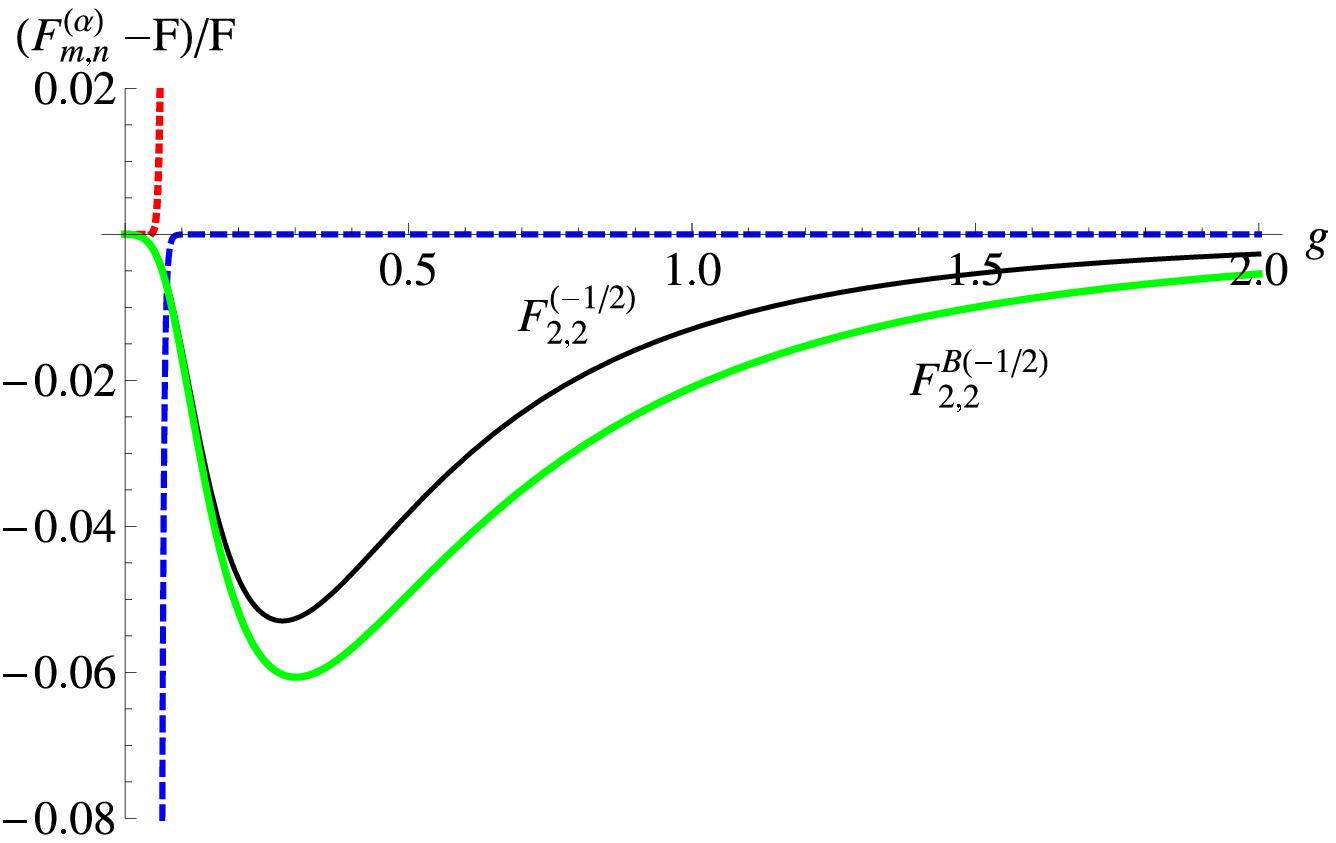}\\
\includegraphics[width=7.4cm]{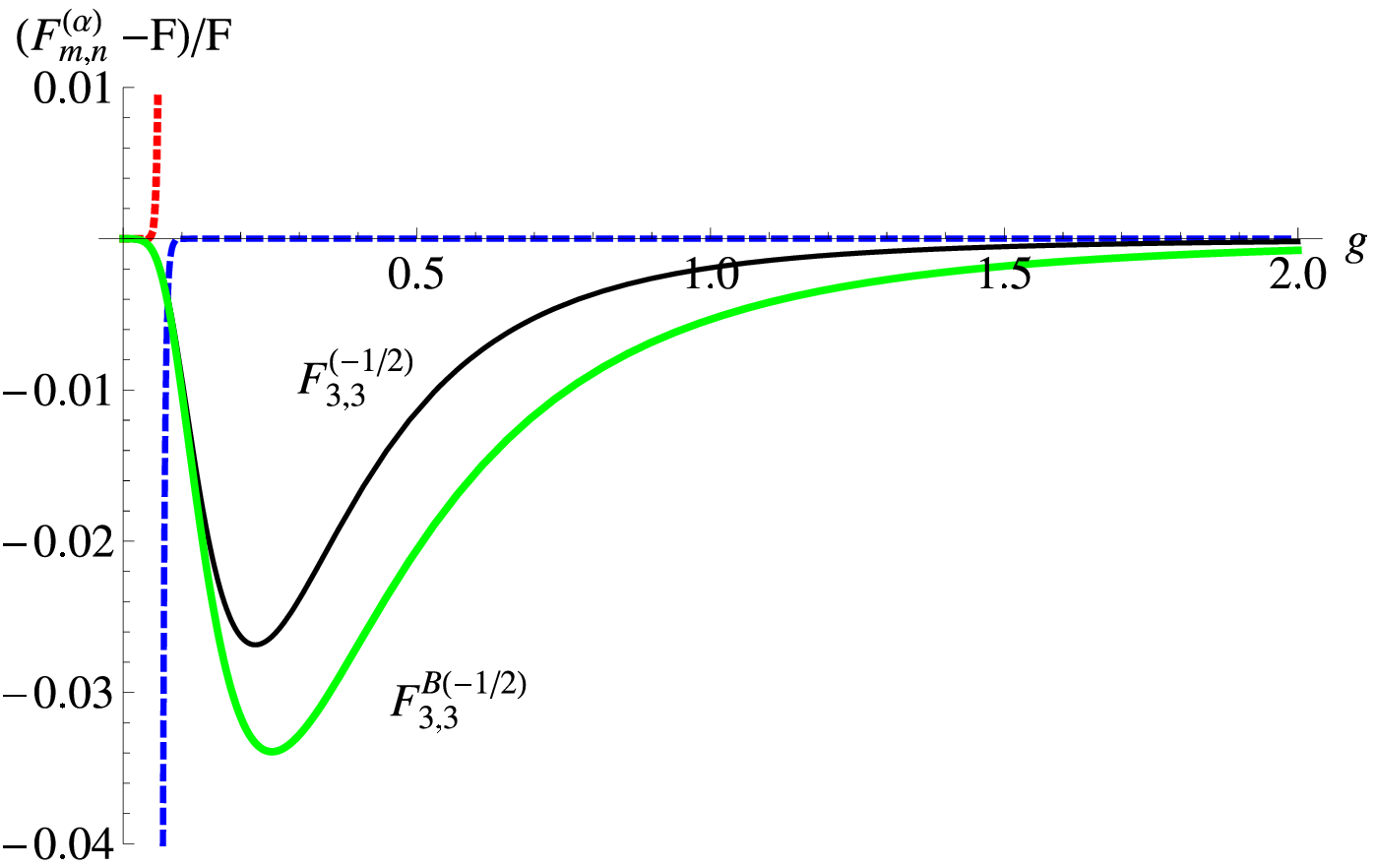}
\includegraphics[width=7.4cm]{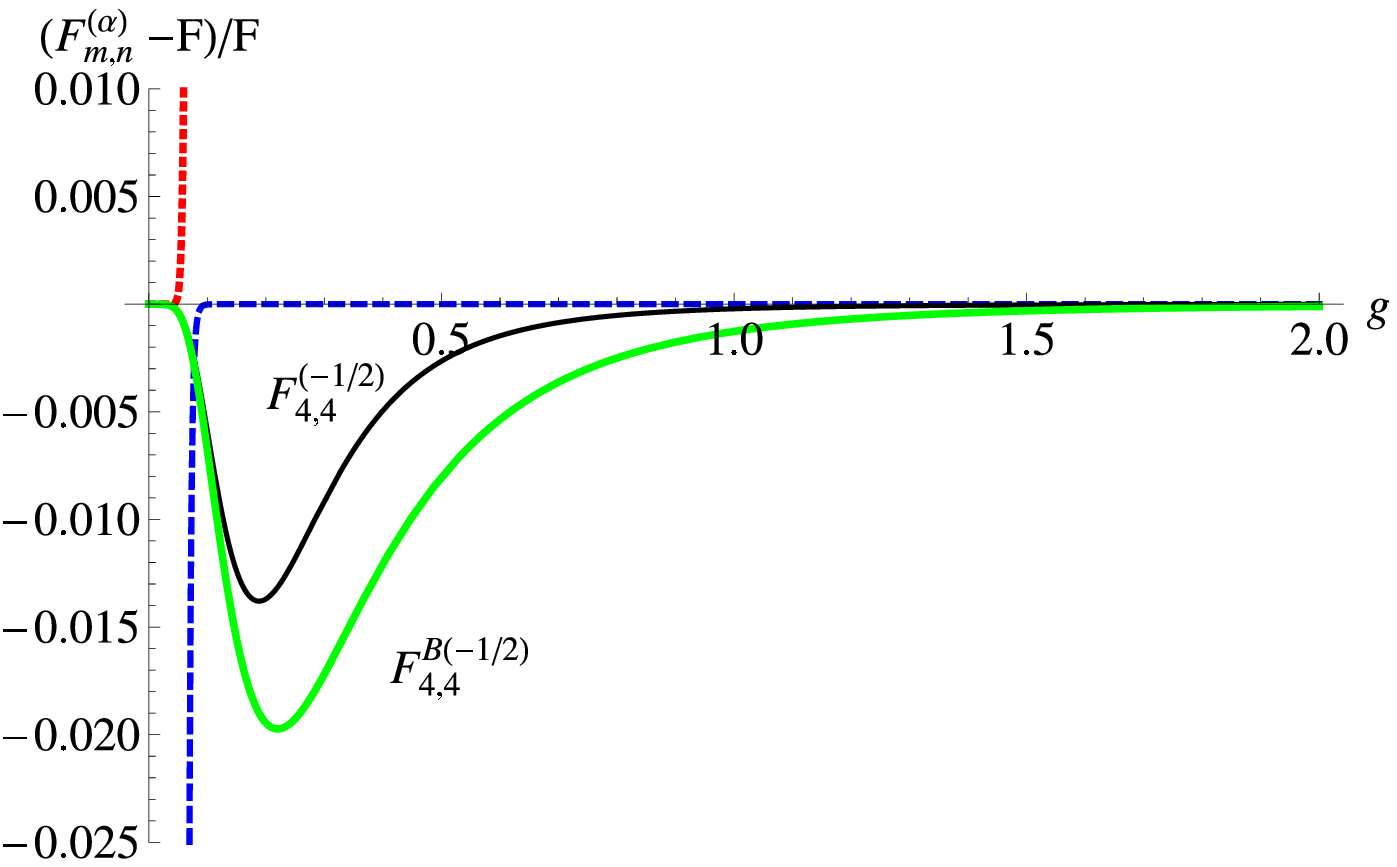}
\end{center}
\caption{Comparison of the Borel FPR $F_{m,n}^{B(\alpha )}$ and usual
  FPR $F_{m,n}^{(\alpha )}$ for $(m,n)=(1,1),(2,2),(3,3),(4,4)$ with
  $(\alpha ,M)=(-1/2 ,10)$.  }
\label{fig:BorelFPR}
\end{figure}

Let us compare the Borel FPR with the usual FPR in the 0d Sine-Gordon
model for ${\rm arg}(g)=0$.  In fig.~\ref{fig:BorelFPR} we plot
\[
\frac{F_s^{(20)}-F}{F}  ,\quad \frac{F_l^{(20)}-F}{F} ,\quad
\frac{F_{m,n}^{(-1/2)}-F}{F} ,\quad \frac{F_{m,n}^{B(-1/2)}-F}{F}  ,
\]
against $g$ for some values of $(m,n)$.  We find that the Borel FPRs
give slightly worse approximations than the usual FPRs with the same
$(m,n,\alpha )$ at least for these four cases.  This differs from our
naive expectation and we have not found any clear reasons for that.
We will not address this issue further, but it would be
interesting to find some good interpretation or modification of our
construction to get better approximation than the usual FPR.

\section{Explicit forms of interpolating functions}
\label{app:explicit}
In this appendix, we write down explicit forms for interpolating
functions used in the main text.
\subsection{Partition function of 0d $\varphi^4$ theory}
\label{app:0dphi4}
\subsubsection{Interpolation along positive real axis of $g$}
\vspace{-2em}
\begin{\eqa}
&& \scriptscriptstyle  
F_{0,0}^{(1/2)}(g) 
=\sqrt{2 \pi }  \left(\frac{8 \pi  g}{   \Gamma \left( 1/4  \right)^2}  +1 \right)^{-1/2} ,\quad
 F_{1,1}^{(1/2)}(g) 
=\sqrt{2 \pi  \Gamma   \left( 1/4  \right) } 
\left( \frac{8 \pi  g \Gamma \left(   1/4 \right)
+\Gamma \left( 1/4 \right)^3     +2 \pi  \Gamma   \left(   -1/4 \right)}
{64 \pi ^2 g^2 +8 \pi  g \Gamma \left( 1/4 \right)^2  +\Gamma   \left( 1/4\right)^4  
+2 \pi  \Gamma \left(-1/4 \right)    \Gamma \left( 1/4\right)} \right)^{1/2} ,\NN\\
&&\scriptscriptstyle F_{1,1}^{(1/6) }(g)
= 2.50663 \left( \frac{1}{6.98929  g^3+ 7.08691 g^2+1}  \right)^{1/6} ,\quad
 F_{2,2}^{(1/2)}(g)
= 2.50663 \sqrt{\frac{37.9117 g^2+10.1532 g +1}   {72.4854 g^3+43.9117 g^2+10.1532 g+1}} ,\NN\\
&&\scriptscriptstyle F_{2,2}^{(1/10)}(g) 
=2.50663 \left( 25.5499 g^5+43.1779 g^4+32.1482 g^3   +30 g^2+1 \right)^{-1/10},   \NN\\
&&\scriptscriptstyle F_{3,3}^{(1/2)}(g) 
=2.50663 \sqrt{\frac{324.019 g^3+110.261 g^2 +16.0304 g+1}
{619.509 g^4+420.201 g^3+116.261 g^2 +16.0304 g+1}} , \NN\\
&&\scriptscriptstyle F_{3,3}^{(1/6)}(g) 
=2.50663\left(\frac{28.2525 g^2+8.0997 g +1}
{  197.465 g^5+256.834 g^4+145.795 g^3  +46.2525 g^2   +8.0997 g+1} \right)^{1/6} ,
\end{\eqa}
\begin{\eqa}
&&\scriptscriptstyle F_{3,3}^{ (1/14)}(g)  
= 2.50663 \left( 93.3994 g^7  +220.976 g^6+239.216 g^5+155.758 g^4+42 g^2+1 \right)^{-1/14} ,\NN\\
&&\scriptscriptstyle F_{4,4}^{(1/2)}(g)
= 2.50663 \sqrt{\frac{3224.56 g^4  +1303.49 g^3+238.239 g^2+22.8745 g+1}
{6165.22 g^5  +4576. g^4  +1440.74 g^3  +244.239  g^2  +22.8745 g+1}} ,\NN\\
&&\scriptscriptstyle F_{4,4}^{(1/10)}(g)
=2.50663 \left[ \frac{ 14.4369 g^2+5.07251 g+1 }
 { 368.86 g^7+752.954 g^6+708.689 g^5  +403.106 g^4  +152.175 g^3  +44.4369 g^2+5.07251 g+1 }  \right]^{1/10} ,\NN\\
&&\scriptscriptstyle F_{4,4}^{(1/18)}(g)  
=2.50663 \left( 341.428 g^9+1038.59 g^8+1475.35 g^7  +1294.34 g^6+780.788 g^5  +594 g^4 +54 g^2+1 \right)^{-1/18} ,\NN\\
&&\scriptscriptstyle F_{5,5}^{(1/2)}(g)  
=1.8128 \sqrt{\frac{(g+0.124648) (g^2+0.133925 g  +0.01439) (g^2+0.203814 g+0.0156882)}{(g^2+0.144222 g+0.0144064)
   (g^2+0.280251 g+0.0278064) (g^2+0.375905 g  +0.0367401)}},\NN\\
&&\scriptscriptstyle F_{6,6}^{(1/2)}(g)  
=1.8128 \sqrt{\frac{(g^2+0.118257 g  +0.0116795) (g^2+0.172869 g+0.0148819) (g^2+0.225907
   g+0.0137177)}{(g+0.179694) ( g^2  +0.118058 g+0.011219) (g^2+0.234356 g+0.0211961) (g^2+0.322914  g+0.0291839)}} ,\NN\\
&&\scriptscriptstyle F_{6,6}^{(1/26)}(g)  
=2.50663 \bigl(4562.58 g^{13}+20047.3 g^{12}+42029.4 g^{11}+55850.9 g^{10}+52692.4 g^9+37452.1 g^8+20758.6 g^7+9156.7 g^6 \NN\\
&&\scriptscriptstyle \ \ \ \ \ \ \ \ \ \ \ 
+3248.63 g^5+929.314 g^4+213.078 g^3+38.4914 g^2+5.28292 g+1 \bigr)^{-\frac{1}{26}}.
\end{\eqa}

\subsubsection{Interpolation along negative real axis of $g$}
\vspace{-2em}
\begin{\eqa}
&& \scriptscriptstyle  
\tilde{F}_{0,0}^{(1/2)}(t) 
=2 \pi  \sqrt{\frac{1}{8 t \Gamma \left(\frac{3}{4}\right)^2+\pi }} ,\quad
\tilde{F}_{1,1}^{(1/2)}(t) 
=2 \pi  \left( \frac{\Gamma \left(\frac{5}{4}\right) \left(8 t \Gamma \left(\frac{3}{4}\right)^2+\pi \right)+2 \Gamma
   \left(\frac{3}{4}\right)^3}{\Gamma \left(\frac{5}{4}\right) \left(64 t^2 \Gamma \left(\frac{3}{4}\right)^4+8 \pi  t \Gamma
   \left(\frac{3}{4}\right)^2+\pi ^2\right)+2 \pi  \Gamma \left(\frac{3}{4}\right)^3} \right)^{1/2} ,\NN\\
&& \scriptscriptstyle  
\tilde{F}_{2,2}^{(1/2)}(t)   
=6.28319 \sqrt{\frac{0.0832 t^2+0.0357 t+0.00906}{t^3+0.0907 t^2+0.1121 t+0.0285}} ,\quad
\tilde{F}_{3,3}^{(1/2)}(t)   
=1.81 \sqrt{\frac{(t+0.398) \left(t^2+0.0631 t+0.0634\right)}{\left(t^2-0.240 t+0.105\right) \left(t^2+0.363 t+0.0626\right)}} ,\NN\\
&& \scriptscriptstyle  
\tilde{F}_{4,4}^{(1/2)}(t)   
=1.81 \sqrt{\frac{t^4+0.40 t^3+0.069 t^2+0.015 t+0.0044}{t^5+0.06 t^4+0.081 t^3+0.011 t^2+0.0040 t+0.00116}} ,\NN\\
&& \scriptscriptstyle  
\tilde{F}_{5,5}^{(1/2)}(t)
=1.81 \sqrt{\frac{(t+0.30) \left(t^2-0.19 t+0.044\right) \left(t^2+0.31 t+0.050\right)}{\left(t^2-0.30 t+0.063\right)
   \left(t^2+0.050 t+0.076\right) \left(t^2+0.34 t+0.037\right)}} ,\NN\\   
&& \scriptscriptstyle  
\tilde{F}_{6,6}^{(1/2)}(t)
=1.8 \sqrt{\frac{(t^2-0.23 t+0.037) (t^2+0.10 t+0.036) (t^2+0.51 t+0.07)}{(t+0.18)
   (t^2-0.30 t+0.050) (t^2-0.042 t+0.07) (t^2+0.20 t+0.042)}} ,\NN\\
&& \scriptscriptstyle  
\tilde{F}_{7,7}^{(1/2)}(t)
=2. \sqrt{\frac{(t+0.3) \left(t^2+0.003 t+0.04\right) \left(t^2+0.4 t+0.05\right)}{\left(t^2-0.1 t+0.06\right) \left(t^2+0.1
   t+0.05\right) \left(t^2+0.4 t+0.03\right)}} ,\NN\\
&& \scriptscriptstyle  
\tilde{F}_{8,8}^{(1/2)}(t)
=1.8 \sqrt{\frac{\left(t^2-0.3 t+0.03\right) \left(t^2-0.04 t+0.03\right) \left(t^2+0.2 t+0.02\right) \left(t^2+0.5
   t+0.06\right)}{(t+0.16) \left(t^2-0.3 t+0.04\right) \left(t^2-0.14 t+0.05\right) \left(t^2+0.09 t+0.05\right)
   \left(t^2+0.2 t+0.02\right)}} ,\NN\\
&& \scriptscriptstyle  
\tilde{F}_{9,9}^{(1/2)}(t)
=1.8 \sqrt{\frac{(t+0.25) \left(t^2-0.25 t+0.026\right) \left(t^2-0.1 t+0.028\right) \left(t^2+0.09 t+0.026\right)
   \left(t^2+0.4 t+0.05\right)}{\left(t^2-0.27 t+0.031\right) \left(t^2-0.16 t+0.04\right) \left(t^2+0.019 t+0.05\right)
   \left(t^2+0.13 t+0.034\right) \left(t^2+0.35 t+0.033\right)}} ,
\end{\eqa}
\subsection{Partition function of 0d Sine-Gordon model}
\label{app:toymodel}

\subsubsection{Interpolation along positive real axis of $g$}
\vspace{-2em}
\begin{\eqa}
&& \scriptscriptstyle  
F_{1,1}^{(-1/2)}(g)
= \sqrt{2} \pi  \left( \frac{2 (2+\pi ) g^2+3 \pi  g+(\pi -1) \pi }{(2+\pi ) g+\pi -1}\right)^{-1/2} ,\quad
F_{2,2}^{(-1/2)}(g)
=\sqrt{2} \pi  \left( \frac{32 (8+5 \pi +5 \pi ^2) g^3+18 \pi  (9+8 \pi ) g^2+3 \pi  (37 \pi -28) g+2 \pi  (10-35
   \pi +16 \pi ^2)}{16 \left(8+5 \pi +5 \pi ^2\right) g^2+\left(-64+41 \pi +32 \pi ^2\right) g+32 \pi ^2-70 \pi +20} \right)^{-1/2} ,\NN\\
&& \scriptscriptstyle  
F_{3,3}^{(-1/2)}(g)
=3.14159 \left( \frac{(g^2-0.0395594 g+0.111122) (g^2+0.828493 g+0.228908)}{(g+0.26163) (g^2+0.0273041   g+0.0618949)} \right)^{-1/2} ,\NN\\
&& \scriptscriptstyle  
F_{4,4}^{(-1/2)}(g)
=3.14159\left( \frac{(g+0.403343) ( g^2-0.10802 g+0.0594072) (g^2+0.474071 g+0.13076)}{\left(g^2-0.0879523
   g+0.0413742\right) \left(g^2+0.357346 g+0.0482101\right)} \right)^{-1/2} ,\NN\\
&& \scriptscriptstyle  
F_{5,5}^{(-1/2)}(g)
=3.14159\left( \frac{(g^2-0.128645 g+0.0352297) (g^2+0.262647 g+0.079582) (g^2+0.623363
   g+0.110622)}{(g+0.189845) \left(g^2-0.127772 g+0.0289318\right) \left(g^2+0.195292 g+0.0359479\right)} \right)^{-1/2} ,\NN\\
&& \scriptscriptstyle  
F_{6,6}^{(-1/2)}(g)
=3.14159 \left( \frac{(g+0.292344) (g^2-0.132606 g+0.0227078) (g^2+0.139599 g+0.0510561) (g^2+0.449886
   g+0.0759508)}{\left(g^2-0.137182 g+0.0208248\right) \left(g^2+0.0850484 g+0.0282403\right) \left(g^2+0.301356
   g+0.0278662\right)} \right)^{-1/2} ,\NN\\
&& \scriptscriptstyle  
F_{7,7}^{(-1/2)}(g)
=3.14159 \left( \frac{(g^2-0.130139 g+0.0157197) (g^2+0.0668157 g+0.0339803) (g^2+0.317179
   g+0.0532544) (g^2+0.489499 g+0.0644541)}{(g+0.148792) \left(g^2-0.134487 g+0.0153154\right) \left(g^2+0.0170188
   g+0.0230983\right) \left(g^2+0.212031 g+0.0221752\right)} \right)^{-1/2} ,\NN\\
 && \scriptscriptstyle  
F_{8,8}^{(-1/2)}(g)
=3.14159\left( \frac{(g+0.228889) (g^2-0.124723 g+0.0115314) (g^2+0.0217909 g+0.0231509) (g^2+0.221648
   g+0.0382247) (g^2+0.391352 g+0.0486489)}{\left(g^2-0.127496 g+0.0115172\right) \left(g^2-0.0219033 g+0.0192889\right)
   \left(g^2+0.136123 g+0.0181046\right) \left(g^2+0.252233 g+0.0179858\right)} \right)^{-1/2} .\NN\\
\end{\eqa}

\subsubsection{Interpolation along negative real axis of $g$}
\vspace{-2em}
\begin{\eqa}
&& \scriptscriptstyle  
\tilde{F}_{1,1}^{(-1/2)}(t) 
=\sqrt{2} \pi \left( \frac{2 (2+\pi ) t^2+3 \pi  t+(\pi -1) \pi }{(2+\pi ) t+\pi -1} \right)^{-1/2} ,\quad
\tilde{F}_{2,2}^{(-1/2)}(t) 
=\sqrt{2} \pi \left( \frac{32 (8+5 \pi +5 \pi ^2) t^3+18 \pi  (9+8 \pi ) t^2+3 \pi  (37 \pi -28) t+2 \pi (10-35 \pi +16 \pi
   ^2)}{16 \left(8+5 \pi +5 \pi ^2\right) t^2+\left(-64+41 \pi +32 \pi ^2\right) t+32 \pi ^2-70 \pi +20} \right)^{-1/2} ,\NN\\
&& \scriptscriptstyle  
\tilde{F}_{3,3}^{(-1/2)}(t) 
=3.14159 \left( \frac{(t^2-0.0395594 t+0.111122) (t^2+0.828493 t+0.228908)}{(t+0.26163) \left(t^2+0.0273041
   t+0.0618949\right)} \right)^{-1/2} ,\NN\\
&& \scriptscriptstyle  
\tilde{F}_{4,4}^{(-1/2)}(t)
= 3.14159 \left( \frac{(t+0.403343) (t^2-0.10802 t+0.0594072) (t^2+0.474071 t+0.13076)}{\left(t^2-0.0879523
   t+0.0413742\right) \left(t^2+0.357346 t+0.0482101\right)} \right)^{-1/2} ,
\end{\eqa}
\begin{\eqa}
&& \scriptscriptstyle  
\tilde{F}_{5,5}^{(-1/2)}(t)
=3.14159\left( \frac{(t^2-0.128645 t+0.0352297) (t^2+0.262647 t+0.079582) (t^2+0.623363
   t+0.110622)}{(t+0.189845) \left(t^2-0.127772 t+0.0289318\right) \left(t^2+0.195292 t+0.0359479\right)} \right)^{-1/2} ,\NN\\
&& \scriptscriptstyle  
\tilde{F}_{6,6}^{(-1/2)}(t)
=3.14159\left( \frac{(t+0.292344) (t^2-0.132606 t+0.0227078) (t^2+0.139599 t+0.0510561) (t^2+0.449886
   t+0.0759508)}{\left(t^2-0.137182 t+0.0208248\right) \left(t^2+0.0850484 t+0.0282403\right) \left(t^2+0.301356 t+0.0278662\right)} \right)^{-1/2} ,\NN\\
&& \scriptscriptstyle  
\tilde{F}_{7,7}^{(-1/2)}(t)
=3.14159\left( \frac{(t^2-0.130139 t+0.0157197) (t^2+0.0668157 t+0.0339803) (t^2+0.317179 t+0.0532544)
   (t^2+0.489499 t+0.0644541)}{(t+0.148792) \left(t^2-0.134487 t+0.0153154\right) \left(t^2+0.0170188 t+0.0230983\right) \left(t^2+0.212031
   t+0.0221752\right)} \right)^{-1/2} ,\NN\\
   && \scriptscriptstyle  
\tilde{F}_{8,8}^{(-1/2)}(t)
=3.14159\left( \frac{(t+0.228889) (t^2-0.124723 t+0.0115314) (t^2+0.0217909 t+0.0231509) (t^2+0.221648
   t+0.0382247) (t^2+0.391352 t+0.0486489)}{\left(t^2-0.127496 t+0.0115172\right) \left(t^2-0.0219033 t+0.0192889\right)
   \left(t^2+0.136123 t+0.0181046\right) \left(t^2+0.252233 t+0.0179858\right)} \right)^{-1/2} .\NN\\
\end{\eqa}

\subsubsection{Borel-FPR}
\vspace{-2em}
\begin{\eqa}
&& \scriptscriptstyle  
\mathcal{B}\tilde{F}_{1,1}^{(-1/2)}(t) 
=\frac{256 \sqrt{\pi }}{2835}
\left( \frac{89260141168201575 t^2+753534959720857600 t+5438758396457123840}{t^{18} (2416615932
   t+10130476945)} \right)^{-1/2} ,\NN\\
&& \scriptscriptstyle  
\mathcal{B}\tilde{F}_{2,2}^{(-1/2)}(t)
= \frac{32 \sqrt{\frac{2 \pi }{11}}}{14175}
\left( \frac{2851044659369509411363841 t^3+21670208776151763289899008
   t^2+93435202549341387371839488 t+454231296657756731387936768}{t^{18} (679261120778364683648 t^2+2276065430590939023520
   t+7445431147866433926325)} \right)^{-\frac{1}{2}} , \NN\\
&& \scriptscriptstyle  
\mathcal{B}\tilde{F}_{3,3}^{(-1/2)}(t)
=0.0000263352\left( \frac{(t^2-1.0468 t+13.0556) (t^2+8.30711 t+23.8912)}{t^{18} (t+2.86868)
   (t^2+0.141631 t+7.48056)} \right)^{-1/2} ,\NN\\
&& \scriptscriptstyle  
\mathcal{B}\tilde{F}_{4,4}^{(-1/2)}(t)
=0.0000263352\left( \frac{(t+4.2183) (t^2-1.833 t+7.96513) (t^2+4.69262 t+14.8922)}{t^{18} (t^2-1.23764
   t+5.48437) (t^2+4.06556 t+6.27685)} \right)^{-1/2} .
\end{\eqa}

\providecommand{\href}[2]{#2}\begingroup\raggedright\endgroup

\end{document}